\documentclass[12pt]{article}
\usepackage{a4wide}
\usepackage{amssymb}
\usepackage{amsmath}
\usepackage{graphicx}
\usepackage{mathdots}
\usepackage{slashed}
\usepackage{verbatim}
\usepackage{xcolor}
\usepackage{setspace}

\usepackage{hyperref}

\usepackage{IEEEtrantools}

\usepackage{collref}

\allowdisplaybreaks

\def\makeatletter{\catcode`\@=11}
\makeatletter
\def\mathbox#1{\hbox{$\m@th#1$}}%
\def\math@ccstyles#1#2#3#4#5#6#7{{\leavevmode
      \setbox0\mathbox{#6#7}%
      \setbox2\mathbox{#4#5}%
      \dimen@ #3%
      \baselineskip\z@\lineskiplimit#1\lineskip\z@
      \vbox{\ialign{##\crcr
             \hfil \kern #2\box2 \hfil\crcr
             \noalign{\kern\dimen@}%
             \hfil\box0\hfil\crcr}}}}
\def\mathaccstyles{\math@ccstyles\maxdimen}
\def\maththroughstyles{\math@ccstyles{-\maxdimen}}
\def\unity%
 {\maththroughstyles{.45\ht0}\z@\displaystyle {\mathchar"006C}\displaystyle 1}

\newcommand{\bZ}{\mathbb{Z}}
\newcommand{\bR}{\mathbb{R}}

\newcommand{\CP}{\mathbb{CP}}

\DeclareMathOperator{\vol}{vol}
\DeclareMathOperator{\arctanh}{arctanh}

\DeclareMathOperator{\re}{Re}
\DeclareMathOperator{\im}{Im}

\begin{document}

\begin{titlepage}

\vskip 2cm

\begin{center}

\begin{spacing}{2}
{\Large \bfseries 
Partition functions for equivariantly twisted $\mathcal{N}=2$ gauge theories on toric K\"ahler manifolds}
\end{spacing}

\vskip 1.2cm

Diego Rodriguez-Gomez\footnote{d.rodriguez.gomez@uniovi.es} and Johannes Schmude\footnote{schmudejohannes@uniovi.es}

\bigskip
\bigskip

\begin{tabular}{c}
Department of Physics, Universidad de Oviedo, \\
Avda.~Calvo Sotelo 18, 33007, Oviedo, Spain
\end{tabular}

\vskip 1.5cm

\textbf{Abstract}
\end{center}

\medskip
\noindent
We consider $\mathcal{N}=2$ supersymmetric pure gauge theories on toric K\"ahler manifolds, with particular emphasis on $\CP^2$. By choosing a vector generating a $U(1)$ action inside the torus of the manifold, we construct equivariantly twisted theories. Then, using localization, we compute their supersymmetric partition functions. As expected, these receive contributions from a classical, a one-loop, and an instanton term. It turns out that the one-loop term is trivial and that the instanton contributions are localized at the fixed points of the $U(1)$. In fact the full partition function can be re-written in a factorized form with contributions from each of the fixed points. The full significance of this is yet to be understood.
\bigskip
\vfill
\end{titlepage}

\setcounter{tocdepth}{2}
\tableofcontents

\section{Introduction}

All the information of a quantum field theory is encoded in the generating functional of its correlation functions. In general this is very hard to compute, yet in some cases and for some other observables such as partition functions and Wilson loops it is possible to perform exact computations that include all non-perturbative effects. For supersymmetric gauge theories in particular and starting with the work of Pestun \cite{Pestun:2007rz}, localization techniques have led to remarkable progress regarding our understanding of such theories in diverse dimensions. Thus, supersymmetric theories emerge as ideal laboratories that allow us to perform exact computations regardless of the strength of the interaction.

At the same time it has become evident that one can obtain a deeper understanding of a supersymmetric gauge theory by placing it on different compact manifolds. See \textit{e.g.}~\cite{Closset:2012ru,Closset:2013vra,Gerchkovitz:2014gta,Gomis:2014woa}. In this paper we take a further step along both of the above directions by considering the partition function of pure $\mathcal{N}=2$ gauge theories with arbitrary classical Lie algebras $\mathfrak{g}$ on generic four-dimensional toric K\"ahler manifolds $M_4$. For the sake of concreteness however, we will illustrate our computations with $M_4=\CP^2$. As the topology of these manifolds is non-trivial, it allows in principle for non-trivial first Chern class for the gauge field. Yet in this paper we will consider the case of vanishing $c_1$.

In general dimension, placing a supersymmetric theory on a compact space in such a way that some supersymmetry is preserved is \textit{per se} a non-trivial task. We will follow the strategy developed in \cite{Festuccia:2011ws,Dumitrescu:2012ha}, which amounts to coupling the gauge theory to supergravity. This way, the supersymmetric couplings to the curved space are automatically fixed. Then, a rigid limit freezes the gravitational dynamics and one is left with the desired supersymmetric gauge theory on the background manifold. An important technical aspect of this approach is that one does not eliminate the auxiliary fields. The supersymmetry algebra closes independently of the equations of motion and the values of the background fields can be found by simply imposing the vanishing of the supersymmetry variations in the supergravity sector.

In the case of Euclidean $\mathcal{N}=2$ theories there are in general two symplectic Majorana Weyl spinors of opposite chirality. There is a degenerate class of solutions for which only spinors of one chirality are used in order to preserve supersymmetry. The Witten (or topological) twist that can be used to define a theory on any four manifold and leads to a topological quantum field theory belongs to this class \cite{Witten:1988ze,Karlhede:1988ax}. We will focus on the general case in which both chiralities are preserved. Here, \cite{Klare:2013dka} showed that the necessary and sufficient condition for supersymmetry is the existence of a conformal Killing spinor $V$ on $M_4$. $V$ is of course a spinor bilinear involving spinors of both chiralities. It plays a crucial role as it twists the superalgebra equivariantly. In our case we will choose $V$ to be a generator of the $\mathbb{T}^2$ torus action on $M_4$.

The topological twist is intimately linked to the computation of the Donaldson invariants of $M_4$, and has thus been studied extensively in the past. See \textit{e.g.}~\cite{Witten:1994ev,Johansen:1994aw,Moore:1997pc,Gottsche:2006tn}. In this paper in turn we are interested in the equivariant version of the theory. As mentioned above, the strategy to compute the equivariant supersymmetric partition function of the pure gage theory on $M_4$ will be to use localization along the lines of \cite{Pestun:2007rz,Gomis:2011pf,Hama:2012bg}. Following what has become by now a fairly standard procedure, we will add a strictly positive $\delta$-exact term $-t\,S_{\rm loc}$ to the action; that is, $S_{\rm loc}=\delta (\int \mathcal{V})$. Here, $\delta$ is any fermionic symmetry of the theory, which in practice one usually chooses to be a combination of BRST and supersymmetry, so that it follows that the partition function does not depend on the parameter $t$. Upon taking the classical limit $t\rightarrow \infty$ the saddle point approximation becomes exact and the partition function is simply given by one-loop fluctuations around the classical action evaluated at the saddle points of $S_{\rm loc}$. One says that the path integral localizes to the localization loci $S_{\rm loc}=0$.

Since K\"ahler manifolds are closely related to Sasakian ones, it is reasonable to expect our theories to have some similarity to the five-dimensional $\mathcal{N}=1$ ones on Sasakian manifolds that were studied in \cite{Hosomichi:2012ek,Kallen:2012va,Qiu:2013pta,Schmude:2014lfa}. Therefore we will occasionally compare the chiral limit of our theories obtained by taking $V$ to zero with the dimensional reduction of the five-dimensional ones. Conversely, it is interesting to wonder to what extend the methods used in this paper can be applied to five dimensional $\mathcal{N}=2$ theories that have so-far been studied in \cite{Qiu:2014cha}.

Returning to the details of our localization calculation, we note that points with non-vanishing $V^2$ define a dense, open subset of $M_4$. Thus there are two types of saddle point configurations. Generic ones for which $V^2 \neq 0$ and a superimposed sector of solutions sitting at the loci where $V^2 = 0$. This is familiar from many, yet not all known examples of localization on four-dimensional manifolds. See \textit{e.g.}~\cite{Assel:2014paa}.
As we will see below, the finite-action configurations\footnote{Infinite action configurations would not contribute to the partition function as their would be weighted by zero.} in the $V = 0$ sector are anti-instantons. We will assume the loci where $V = 0$ to be isolated which corresponds to a slight restriction on the possible choices of $V$ in $\mathbb{T}^2$. Close to any such point the background takes the form of a copy of the $\Omega$ background \cite{Nekrasov:2002qd,Nekrasov:2003rj} and we can compute the instanton contribution to the partition function by appropriately gluing copies of the Nekrasov instanton partition function.

Somewhat remarkably, we find that the fluctuation determinant for the case of  toric K\"ahler manifolds considered is simply $1$. Since in addition the classical action can also be written as a sum of three copies of a function evaluated at precisely the $\Omega$ background parameters corresponding to the loci $V^2=0$, we immediately find an interesting factorization of the partition function whose implications remain yet to be fully understood.

The structure of this paper is as follows: Section \ref{sec:SUSY} begins with a summary of the relevant parts of the $\mathcal{N}=2$ conformal supergravity that are needed for the subsequent construction of the gauge theory. Studying the BPS equations arising from the gravitino and dilatino variations we then find the supersymmetric background and as well as the Killing spinors for both the topological and the equivariant twists. This allows us to define the gauge theory on the curved space $M_4$. We also construct the cohomological version of the supersymmetry algebra, which explicitly shows the equivariant twist. In section \ref{sec:localization} we study the localization locus of the gauge theory. To begin, we do so by directly studying the BPS equations of the vector multiplet. In section \ref{sec:localizing_action} we write down an explicitly localization term $S_{\rm loc}$ and show that the solution to the BPS equations precisely coincides with the set of configurations on which $S_{\rm loc}=0$. While these configurations correspond to $V^2\ne 0$, we study the instanton solutions sitting at the locus $V^2=0$in section \ref{sec:instantons}. Finally, we discuss the gauge fixing sector, which combines the BRST with the SUSY complex (and plays an interesting subtle role in fully determining the relevant localization locus. See below.) Then, in section \ref{sec:partition_function} we compute the partition function by explicitly writing down the classical, one-loop and instanton contributions. Remarkably, these three contributions can be written in a factorized form as the product of a function evaluated at the $\Omega$ backgrounds around each of the points where $V^2=0$. We end with some conclusions and future projects in section \ref{conclusions}. We leave for the appendices technical aspects of relevance for the computations in the main text.

\vspace{.5cm}

\textbf{Note added}: In the closing stages of this project we received  \cite{Bawane:2014uka,Sinamuli:2014lma}, which consider $M_4=\mathbb{P}^1\times \mathbb{P}^1$. Especially \cite{Bawane:2014uka} obtains, where applicable, similar results to ours.

\section{Rigid supersymmetry on toric K\"ahler manifolds}\label{sec:SUSY}

Our aim is to study $\mathcal{N}=2$ gauge theories on toric K\"ahler manifolds, with $\CP^2$ being our star example. Hence, our first task will be the construction of the supersymmetric lagrangian of the theory on the curved space. Following \cite{Festuccia:2011ws}, we couple the gauge theory to supergravity and then take a rigid limit so that the globally supersymmetric lagrangian automatically emerges. Following \cite{Klare:2013dka}, we will use four-dimensional $\mathcal{N}=2$ conformal supergravity and refer to \cite{vanProyen,Freedman:2012zz} for a thorough introduction to the subject. The field content of the Weyl multiplet is
\begin{equation}
  g_{mn}\,, \quad
  D\,, \quad
  T_{mn}\,, \quad
  A_x\,, \quad
  A_4\,, \quad
  \psi_\pm^i\,, \quad
  \chi_\pm^i\,.
\end{equation}
Here $A_x$ and $A_4$ are the connections for the $SU(2)$ and $U(1)$ R-symmetries, $T_{mn}$ is a two-form, and $D$ a scalar. Furthermore, $x$ and $i$ are adjoint and fundamental $SU(2)$ indices respectively. After Wick rotation \cite{vanNieuwenhuizen:1996tv}, the supersymmetry transformations of the fermions are
\begin{subequations}\label{eq:SUSY_variations_Weyl-multiplet}
\begin{IEEEeqnarray}{rCl}
    \delta \psi^i_{+m} &=& \nabla_m \epsilon^i_+ + \frac{\imath}{2}
    A_{mx} \sigma^{xi}_{\phantom{xi}j} \epsilon^j_+ + \frac{1}{2}
    A_{m4} \epsilon^i_+ + \frac{\imath}{4} T^+_{mn} \gamma^n
    \epsilon^i_- - \gamma_m \eta^i_-\,, \\
    \delta \psi_{-m}^i &=& \nabla_m \epsilon^i_- + \frac{\imath}{2}
    A_{mx} \sigma^{xi}_{\phantom{xi}j} \epsilon^j_-  - \frac{1}{2}
    A_{m4} \epsilon^i_- + \frac{\imath}{4} T^-_{mn} \gamma^n
    \epsilon_+^i - \gamma_m \eta^i_+\,, \\
    \delta \chi^i_+ &=& \frac{\imath}{6} (\nabla^m + A^{m4}) T^+_{mn}
    \gamma^n \epsilon^i_- - \frac{1}{6} dA_4 \cdot \gamma \epsilon_+^i
    + \frac{D}{3} \epsilon^i_+ + \frac{\imath}{12} \gamma \cdot T^+
    \eta^i_+ \nonumber\\ 
    &&+ \frac{\imath}{6} \left( \partial_{[m} A_{n]}^x + \frac{1}{2}
      A_m^y A_n^z \epsilon^{yzx} \right) \gamma^{mn}
    \sigma^{xi}_{\phantom{xi}j} \epsilon^j_+\,, \\ 
    \delta \chi^i_- &=& \frac{\imath}{6} (\nabla^m - A^{m4}) T^-_{mn}
    \gamma^n \epsilon^i_+ + \frac{1}{6} dA_4 \cdot \gamma \epsilon^i_-
    + \frac{D}{3} \epsilon^i_- + \frac{\imath}{12} \gamma \cdot T^-
    \eta^i_- \nonumber\\ 
    && + \frac{\imath}{6} \left( \partial_{[m} A_{n]}^x + \frac{1}{2}
      A_m^y A_n^z \epsilon^{yzx} \right) \gamma^{mn}
    \sigma^{xi}_{\phantom{xi}j} \epsilon^j_-\,.
    \end{IEEEeqnarray}
\end{subequations}

In addition to the Weyl multiplet, we consider a vector multiplet. Eventually and upon taking the rigid limit, the field theory of interest will be that of this vector multiplet. The standard $\mathcal{N}=2$ vector multiplet contains a complex scalar $\phi$, an auxiliary $SU(2)$ triplet $Y_{ij}$, the gauge connection $\mathcal{A}$ and the gaugino. Note that, after Wick rotation, $\phi$ and $\bar{\phi}$ are a priori independent. The Wick-rotated supersymmetry variations for the vector multiplet are
\begin{subequations}\label{eq:SUSY_variations_vector-multiplet}
\begin{IEEEeqnarray}{rCl}
    \delta \phi^I &=& -\frac{\imath}{2} \epsilon_+^i B \Omega^I_{i+}\,, \\
    \delta \bar{\phi}^I &=& \frac{\imath}{2} \epsilon_{-i} B \Omega^{Ii}_-\,, \\
    \delta \mathcal{A}^I_m &=& \frac{1}{2} \epsilon^{ij} \epsilon_{i-} B \gamma_m
    \Omega^I_{j+} + \frac{1}{2} \epsilon_{ij} \epsilon_+^i B \gamma_m
    \Omega^{Ij}_-\,, \\
    \delta \Omega^{Ii}_+ &=& \imath \slashed{D} \phi^I \epsilon^i_- -
    \frac{1}{4} \gamma^{ab} \left( F_{ab}^{I+} - \frac{1}{2}
      \bar{\phi}^I T^+_{ab} \right) \epsilon^i_+ + \frac{1}{2}
    Y^{Ii}_{\phantom{Ii}j} \epsilon^j_+ - g \phi^J \bar{\phi}^K
    f_{JK}^{\phantom{JK}I} \epsilon^i_+ \nonumber \\  
    && + 2\imath \phi^I \eta_+^i - g \alpha^J \Omega_+^{Ki} f_{JK}^{\phantom{JK}I}\,, \\
    \delta \Omega^{Ii}_- &= &-\imath \slashed{D} \bar{\phi}^I
    \epsilon_+^i + \frac{1}{4} \gamma^{ab} \left( F^{I-}_{ab} -
      \frac{1}{2} \phi^I T^-_{ab} \right) \epsilon^i_- - \frac{1}{2}
    Y^{Ii}_{\phantom{Ii}j} \epsilon^j_- - g \phi^J \bar{\phi}^K
    f_{JK}^{\phantom{JK}I} \epsilon^i_- \nonumber \\  
    && -2\imath \bar{\phi}^I \eta_-^i - g \alpha^J \Omega_-^{Ki} f_{JK}^{\phantom{JK}I}\,, \\
    \delta Y_{ij}^I &=& \epsilon_{(i-} B \slashed{D} \Omega^I_{j)+} +
    \epsilon_{ik} \epsilon_{jl} \epsilon^{(k}_+ B \slashed{D}
    \Omega^{l)I}_- 
    + 2 \imath g \epsilon_{k(i} \left(\epsilon_{j)-} B
      \phi^J \Omega^{kK}_- + \epsilon^k_+ B \bar{\phi}^J
      \Omega^K_{j)+} \right) f_{JK}^{\phantom{JK}I}\,.
\end{IEEEeqnarray}
\end{subequations}
Here, the covariant derivatives appearing in the supersymmetry transformations are
\begin{equation}\label{eq:covariant-derivatives_defined}
  \begin{aligned}
    D_m \Omega^{Ii}_+ &= \nabla_m \Omega^{Ii}_+ + \frac{\imath}{2} A_{mx} \sigma^{xi}_{\phantom{xi}j} \Omega^{Ij}_+ + g [\mathcal{A}_m, \Omega^{i}_+]^I\,, \\
    D_m \Omega^{Ii}_- &= \nabla_m \Omega^{Ii}_- + \frac{\imath}{2} A_{mx} \sigma^{xi}_{\phantom{xi}j} \Omega^{Ij}_- + g [\mathcal{A}_m, \Omega^{i}_-]^I\,, \\
    D_m \phi^I &= \partial_m \phi^I + g [W_m, \phi]^I\,.
  \end{aligned}
\end{equation}
These transformations leave the action of the gauge theory invariant, which can be taken from \cite{Klare:2013dka}. Its bosonic part is
\begin{equation}\label{eq:vector_multiplet_action}
  \begin{aligned}
    \mathcal{L} &= d \phi \bar{\phi} + \nabla^A_m \phi \nabla^{Am} \bar{\phi} + \frac{1}{8} Y^i_{\phantom{i}j} Y^j_{\phantom{j}i} - g [\phi, \bar{\phi}]^2 + \frac{1}{8} F_{mn} F^{mn} \\
    &- \frac{1}{4} (\phi F_{mn} T^{+mn} + \bar{\phi} F_{mn} T^{-mn}) - \frac{1}{16} (\phi^2 T^+_{mn} T^{+mn} + \bar{\phi}^2 T^-_{mn} T^{-mn} )\,.
  \end{aligned}
\end{equation}
Up to conventions, this agrees with the action of \cite{Hama:2012bg}.

\subsection{Supersymmetric backgrounds}

Since the super Yang-Mills theory on the curved space arises from the rigid limit of the combined supergravity plus vector multiplet system, the relevant background for the later can be found by imposing the vanishing of the Weyl multiplet supersymmetry variations in eqs.~\eqref{eq:SUSY_variations_Weyl-multiplet}. Solving these fully determines the supersymmetry variations of the vector multiplet \eqref{eq:SUSY_variations_vector-multiplet} as well as the action \eqref{eq:vector_multiplet_action}. 

In order to provide a very explicit example, we will first construct the Killing spinors for $\CP^2$ before generalizing to arbitrary toric K\"ahler manifolds.

\subsubsection{$M_4=\CP^2$}

For $\CP^2$ we use the metric
\begin{equation}
\label{eq:explicitmetric}
ds^2=d\rho^2+\frac{\sin^2\rho}{4}\,\Big[ \sigma_1^2+\sigma_2^2+\cos^2\rho\,\sigma_3^2\Big]\,,
\end{equation}
with Maurer-Cartan forms
\begin{equation}
\sigma_1=\cos\psi\,d\theta+\sin\psi\,\sin\theta\,d\phi\,,\qquad \sigma_2=\sin\psi\,d\theta-\cos\psi\,\sin\theta\,d\phi\,, \qquad \sigma_3=d\psi+\cos\theta\,d\phi\,,
\end{equation}
and $\rho \in [0,\pi/2]$, $\theta \in [0, \pi]$, $\phi \in [0, 2\pi]$, and $\psi \in [0, 4\pi]$. The two torus is generated by the Killing vectors $\partial_\phi$, $\partial_\psi$. We choose the frame
\begin{equation}
e^1=d\rho\,,\qquad e^2=\frac{\sin\rho\,\cos\rho}{2}\,\sigma_3\,,\qquad e^3=\frac{\sin\rho}{2}\,\sigma_1\qquad e^4=\frac{\sin\rho}{2}\,\sigma_2,.
\end{equation}

Defining
\begin{equation}
z_1=\tan\rho\,\cos\frac{\theta}{2}\,e^{i\,\frac{\psi+\phi}{2}}\,,\qquad z_2=\tan\rho\,\sin\frac{\theta}{2}\,e^{i\,\frac{\psi-\phi}{2}}\,,
\end{equation}
the metric can be rewritten in terms of the K\"ahler potential $K=\log(1+|z_1|^2+|z_2|^2)$,
\begin{equation}
ds^2=\frac{\partial^2\,K}{\partial\,z_i\,\partial\,\bar{z}_j}\,dz_i\,d\bar{z}_j\,.
\end{equation}
Furthermore
\begin{equation}
  J=\frac{i}{2} \partial \bar{\partial}K = e^1 \wedge e^2 + e^3 \wedge e^4 = \frac{1}{2} d\Theta,
  \qquad
  \Theta=\frac{\sin^2\rho}{2} \sigma_3\,.
\end{equation}

After calculating the spin connection, $de^a + \omega^a_{\phantom{a}b} e^b = 0$, one finds two negative chirality spinors $\epsilon_-^i$ satisfying $\partial_m \epsilon_-^i = 0$ as well as the projections $\gamma^{12}\epsilon_-^i = \gamma^{34} \epsilon_-^i = \imath \sigma^{3i}_{\phantom{3i}j} \epsilon_-^j$. Their Killing spinor equation is
\begin{equation}\label{eq:Killing_spinor_equation}
  \nabla_m \epsilon_-^i - \frac{3\imath}{2} \Theta_m \sigma^{3i}_{\phantom{3i}j} \epsilon_-^j = 0\,.
\end{equation}

Comparing \eqref{eq:Killing_spinor_equation} with the SUSY variations \eqref{eq:SUSY_variations_Weyl-multiplet}, one sees that $\delta \psi^i_{+m} = \delta \psi^i_{-m} = 0$ if
\begin{equation}
  A_3 = -3 \Theta\,, \qquad
  A_4 = T^+ = \epsilon_+^i = 0\,.
\end{equation}
A similar analysis for the dilatino variations $\delta \chi^i_\pm$ imposes $D = 6$. One can verify this using the equations in \cite{Klare:2013dka}. Due to a difference in notation, the above $D = 6$ corresponds to $d = 0$ in that paper. Note that this causes the $\phi\,\bar{\phi}$ mass-like term in \eqref{eq:vector_multiplet_action} to vanish, as opposed to the case of squashed spheres.

The solution which we have found involves only negative chirality spinors. In fact, it just corresponds to the familiar topologically twisted theory. In order to construct the equivariantly twisted theory we need to add positive chirality spinors, so that we can construct a vector-like spinor bilinear providing the equivariant parameters. To add positive chirality spinors, we pick a generic Killing vector $V$ generating a $U(1)$ action inside the torus. We can parametrize it as $V = p_\psi \partial_\psi + p_\phi \partial_\phi$ for $p_\psi, p_\phi \in \bR$. As we will see below, these $p_{\psi},\,p_{\phi}$ are essentially the equivariant parameters. Note that 

\begin{equation}
\label{eq:localV2}
V^2=\frac{1}{4}\,\Big(\,p_{\phi}^2\,\sin^2\theta\,\sin^2\rho+(\frac{p_{\psi}+p_{\phi}\,\cos\theta}{2})^2\,\sin^22\rho\Big)\,.
\end{equation}
Hence, for generic $p_{\psi},\,p_{\phi}$, $V^2$ vanishes at $\rho=0$, $\{\rho=\frac{\pi}{2},\,\theta=0\}$ and $\{\rho=\frac{\pi}{2},\,\theta=\pi\}$. Note however that, for particular choices of $p_{\psi}$ and  $p_{\phi}$, $V^2$ vanishes on more generic subspaces.\footnote{For example , if $p_{\psi}=p_{\phi}$, then $V^2$ vanishes at $\theta=\pi$ for any value of $\rho$. Another example is $p_{\psi}=0$ or $p_{\phi}=0$, when we find that $V^2$ vanishes for $\rho=\{0,\,\frac{\pi}{2}\}$ regardless of $\theta$.} In the following we will assume that $p_{\psi},\,p_{\phi}$ take generic values in such a way that $V^2=0$ only happens at the three reported points.

With this $V$ we can construct positive chirality spinors as $\epsilon_+^i = \imath \slashed{V} \epsilon_-^i$. A direct analysis of the gravitino equations imposes
\begin{equation}
\label{eq:graviphoton}
  \begin{aligned}
    T^- &= 0\,, &\qquad
    T^+ &= -2 dV^+\,.
  \end{aligned}
\end{equation}
As in the previous case, the dilatino variations vanish for $D = 6$ or $d = 0$ respectively.

\subsubsection{Toric K\"ahler manifolds}\label{sec:toric_kahler}

In this section, we generalize the results of the previous section to any four-dimensional toric K\"ahler manifold $M_4$. Such manifolds can be defined as closed connected $4$-dimensional K\"ahler manifolds with an effective Hamiltonian holomorphic action of the real $2$-torus $\mathbb{T}^2$. However, for our purposes it is best to think of the Delzant construction \cite{delzant1988hamiltoniens} and the work of Guillemin and Abreu \cite{guillemin1994kaehler,abreu2003kahler}, which we will quickly review here. Further details are in appendix \ref{sec:appendix_abreu}. To start, one introduces symplectic coordinates $(x^i, y_i)$, $i = 1, 2$ with the $y_i$ parametrizing the $\mathbb{T}^2$ and the $x_i$ being the coordinates of the Delzant polytope $P$. The most familiar example is $\CP^2$ with the polytope defined by $0 \leq x^1$, $0 \leq x^2$, and $x^1 + x^2 \leq 1$. For $\CP^1 \times \CP^1$, one has $0 \leq x^{1,2} \leq 1$. On each edge of the polytope the torus collapses to an $S^1$. Thus, the vertices are the fixed points of the torus action. The symplectic form is $\omega = dx^i \wedge dy_i$ and metric and almost complex structure are given in terms of a potential function $g(x) = g_P(x) + h(x)$. Assume the polytope is defined by inequalities $\langle x, \mu_r \rangle \geq \lambda_r$, $r = 1, \dots, d$, each $\mu_r$ being a primitive element of the lattice $\bZ^2 \subset \bR^2$ and inward pointing normal to the $r$-th $(n-1)$-dimensional face of $P$. Then, the canonical potential $g_P(x)$ is defined in terms of the functions $l_r : \bR^2 \to \bR$, defined by
\begin{equation}
  l_r (x) = \langle x, \mu_r \rangle - \lambda_r
\end{equation}
as
\begin{equation}
  g_P(x) = \frac{1}{2} \sum_{r=1}^d l_r(x) \log l_r(x)\,.
\end{equation}
Define $G = \text{Hess}_x (g)$, i.e.~$(G)_{ij} = \partial_{x^i} \partial_{x^j} g$. Then
\begin{equation}
  J = \begin{pmatrix}
    0 & - G^{-1} \\ G & 0
  \end{pmatrix}\,,
  \qquad
  ds^2 = \begin{pmatrix}
    G & 0 \\ 0 & G^{-1}
  \end{pmatrix}.
\end{equation}
The function $h(x)$ has to be smooth on $P$ and chosen such that there is a smooth and strictly positive function $\delta(x)$ satisfying
\begin{equation}
  \det G = \left[ \delta(x) \prod_{r=1}^d l_r(x) \right]^{-1}\,.
\end{equation}

Any K\"ahler manifold $M_4$ admits a spinor $\psi$ satisfying\footnote{
  For details we refer to the summary in \cite{Martelli:2006yb} and the references therein. In the conventions of \cite{Martelli:2006yb}, $\psi$ is the constant section of $\bigwedge^{0,\text{even}}T_M^* \cong \mathcal{V}^+$; i.e.~has positive chirality.
}
\begin{equation}\label{eq:Kahler_spinor_equation}
  \nabla_Y \psi = \frac{\imath}{2} A_{\text{Ric}}(Y) \psi
\end{equation}
with the connection one-form $A_{\text{Ric}}$ defined by $dA_{\text{Ric}} = \rho$ where $\rho$ is the Ricci form of $M$. Here $\rho$ is defined in terms of the Ricci tensor and the complex structure as $\rho (X, Y) = \text{Ric}(JX, Y)$ \cite{huybrechts2005complex}. The symplectic Majorana conjugate of \eqref{eq:Kahler_spinor_equation} satisfies $\nabla_Y \psi^* = -\frac{\imath}{2} A_{\text{Ric}}(Y) \psi^*$. To match this with our calculation for $\CP^2$, we note that $\CP^2$ carries an Einstein metric. Thus $\text{Ric} = 6g$ and $\rho = -6 J$. With $d\Theta = 2J$, one sees that $A_{\text{Ric}} = -3\Theta$.

By comparison with our previous results it is clear that
\begin{equation}
  \epsilon_-^1 = \psi^*\,, \qquad
  \epsilon_-^2 = \imath B \psi\,, \qquad
  A_3 = A_{\text{Ric}}
\end{equation}
solve the gravitino and dilatino equations in the absence of $\epsilon_+^i$. Just as before, this corresponds to the topological twist. In order to construct the equivariantly twisted theory, we pick a generic Killing vector
\begin{equation}
  V = p\, \partial_{y_1} + q\, \partial_{y_2}, \qquad p, q \in \mathbb{R}\,,
\end{equation}
and define $\epsilon_+^i = \imath \slashed{V} \epsilon_-^i$. Since $\nabla_\mu^A \epsilon_-^i = 0$, we have $\nabla_\mu^A \epsilon_+^i = \frac{\imath}{2} dV^+_{\mu\nu} \gamma^\nu \epsilon_-^i$ and the gravitino equations are solved by $T^+ = -2 dV^+$ and $T^- = 0$. Again, one fixes the scalar fields $D$ or $d$ by solving the dilatino variation. And once again, one finds $d = 0$ in the notation of \cite{Klare:2013dka} meaning that the mass term in \eqref{eq:vector_multiplet_action} vanishes. If we choose a vielbein such that $J = e^{12} + e^{34}$, we can maintain the projections for $\epsilon_-^i$.

Generalizing the $\CP^2$ case, we restrict $p,\,q$ such that $V^2$ vanishes only at certain isolated points in the manifold. One can see -- \textit{c.f.}~appendix \ref{sec:appendix_abreu} -- that these correspond to the vertices of the Delzant polytope -- of which there were three in the above discussion of $\CP^2$. Nevertheless, exactly as in the $\CP^2$ case and for certain choices of $p$ and $q$, $V^2$ can vanish at more generic loci, namely $\CP^1$s corresponding to edges of the polytope.

\subsection{Cohomological form of the supersymmetry transformations}
\label{sec:the_cohomological_complex}

Substituting the background fields as well as the Killing spinors from the previous sections into eq.~\eqref{eq:vector_multiplet_action} gives us the lagrangian for the gauge theory on toric K\"ahler manifolds. In turn, the supersymmetry variations can be found from eqs.~\eqref{eq:SUSY_variations_vector-multiplet}. 

We now bring the supersymmetry transformations into standard cohomological form. Details are relegated to appendix \ref{sec:appendix_cohomological_susy_technicalities}. To begin, we note that since $\eta_-^i = \frac{\imath}{8} dV_{ab} \gamma^{ab} \epsilon_-^i$ and $T^+ = -2 dV^+$, we can rewrite the gaugino variations \eqref{eq:SUSY_variations_vector-multiplet} as
\begin{equation}
  \begin{aligned}
    \delta \Omega^i_+ &= \imath \slashed{D} \phi \epsilon_-^i - \frac{1}{4} (F^+_{ab} - \frac{1}{2} \bar{\phi} T^+_{ab}) \gamma^{ab} \epsilon_+^i + \frac{1}{2} Y^i_{\phantom{i}j} \epsilon_+^j - g [\phi, \bar{\phi}] \epsilon_+^i\,, \\
    \delta \Omega^i_- &= - \imath \slashed{D} \bar{\phi} \epsilon_+^i + \frac{1}{4} (F^-_{ab} - \frac{1}{2} \bar{\phi} T^-_{ab}) \gamma^{ab} \epsilon_-^i - \frac{1}{2} Y^i_{\phantom{i}j} \epsilon_-^j - g [\phi, \bar{\phi}] \epsilon_-^i\,,
  \end{aligned}
\end{equation}
and without any $\eta_\pm^i$ terms yet with $T = -2 dV$. We define $\mathcal{F} = F - \frac{1}{2} \bar{\phi} T$.

Next, we define Grassmann odd forms $\eta \in \Omega^0$, $\Psi \in \Omega^1$, and $\chi \in \Omega^- \subset \Omega^2$.
\begin{equation}
  \begin{aligned}
    \chi &= \epsilon_{ij} \epsilon^i_- B \gamma_{(2)} \Omega^j_-\,, \\
    \Psi &= \frac{1}{2} \left( \epsilon^{ij} \epsilon_{-i} B \gamma_{(1)} \Omega_{+j} + \epsilon_{ij} \epsilon_+^i B \gamma_{(1)} \Omega_-^j \right)\,, \\
    \eta &= -\frac{\imath}{2} \epsilon_{ij} \epsilon_{-}^i B \Omega_{-}^j\,.
  \end{aligned}
\end{equation}
These definitions are invertible. Concerning the bosonic modes, we rewrite the $SU(2)$ triplet $Y_{ij}$ in terms of an anti self-dual two form,
\begin{equation}
  H = - 2 \imath \mathcal{F}^- + \frac{\imath}{2} \mathcal{M}^{ij}_- Y_{ij} - 4 \imath (D\bar{\phi} \wedge V)^-\,.
\end{equation}
Here, $\mathcal{M}_-^{ij} = - \imath \epsilon_-^i B \gamma_{(2)} \epsilon_-^j$ as in \eqref{eq:M-triplets} in appendix \ref{sec:conventions}. The definition of $H$ is such that $\delta \chi = H$. In terms of the variables $\mathcal{A}, \phi, \bar{\phi}, H$ and $\eta, \Psi, \chi$, the algebra is
\begin{equation}\label{eq:cohomological_SUSY_algebra}
  \begin{aligned}
    \delta \bar{\phi} &= \eta\,, &\qquad
    \delta \eta &= \pounds_V \bar{\phi} + G_{\phi - V^2 \bar{\phi} - \imath_V \mathcal{A}}[\bar{\phi}]\,, \\
    \delta \mathcal{A} &= \Psi\,, &\qquad
    \delta \Psi &= \pounds_V \mathcal{A} + G_{\phi - V^2 \bar{\phi} - \imath_V \mathcal{A}}[\mathcal{A}], \\
    \delta \chi &= H\,, &\qquad
    \delta H &= \pounds_V \chi + G_{\phi - V^2 \bar{\phi} - \imath_V \mathcal{A}}[\chi]\,,
  \end{aligned}
\end{equation}
with $G_\theta$ denoting gauge transformations and defined in \eqref{eq:Yang-Mills_conventions}. See equations \eqref{eq:cohomological_susy_variations_appendix} for a formulation of the above that will be useful when solving the BPS equations in the next section. In the form of \eqref{eq:cohomological_SUSY_algebra} it is clear that we have a complex
\begin{equation}\label{eq:cohomological_complex}
  \begin{aligned}
    Z &\in \{ \bar{\phi}\,, \mathcal{A}, \chi \}, &\qquad
    Z^\prime &\in \{ \eta, \Psi, H \}\,, \\
    \delta Z  &= Z^\prime\,, &\qquad
    \delta Z^\prime &= \pounds_V Z + G_{\phi - V^2 \bar{\phi} - \imath_V \mathcal{A}} [Z]\,,
  \end{aligned}
\end{equation}
with $\mathcal{A}, \Psi \in \Omega^1$, $H, \chi \in \Omega^-$, and $ \bar{\phi}, \eta \in \Omega^0$. This is essentially the equivariant complex of \cite{Pestun:2007rz,Qiu:2013pta,Kallen:2011ny,Kallen:2012cs}. Per usual, one of the scalars -- here $\phi$ -- is somewhat special:
\begin{equation}
  \begin{aligned}
    \delta \phi &= \imath_V \Psi + V^2 \eta\,, &\qquad
    \delta^2 \phi &= \pounds_V \phi + G_{\phi - V^2 \bar{\phi} - \imath_V \mathcal{A}}[\phi]\,.
  \end{aligned}
\end{equation}
Thus $\delta \phi = \delta (\imath_V \mathcal{A} + V^2 \bar{\phi})$ and the gauge-parameter $\phi - V^2 \bar{\phi} - \imath_V \mathcal{A}$ is invariant under supersymmetry transformations. Furthermore, note that the gauge parameter has an immediate dependence on $V^2$, the norm of the equivariant vector.

\section{Localization}\label{sec:localization}

Having defined supersymmetric gauge theories on toric K\"ahler manifolds, we are now interested in their supersymmetric partition functions, which we will compute using localization. As it is customary, we deform the action with a $\delta$-exact term $-t\, S_{\rm loc}$. This introduces $t^{-1}$ as a new effective $\hbar$ on which the partition function does not depend. Then, upon taking the classical limit $t\rightarrow \infty$, the saddle point approximation becomes exact, and the partition function is simply given by the product of the classical action evaluated at the saddle points of the localization action times the fluctuation determinant. Hence, our first task will be to study this localization locus.

In the following we will concentrate on the $\CP^2$ case. Nevertheless, the results hold in the case of generic toric K\"ahler manifolds upon performing the obvious substitutions.

\subsection{The localization locus}

We start by finding the localization locus on which the partition function localizes. Since these correspond to supersymmetric configurations, we can as well derive them by studying the BPS equations. In section \ref{sec:localizing_action} we will consider the explicit form of the $\delta$-exact localization term $S_{\text{loc}}$ that is be added to the action to localize the path integral and show that the configurations arising from the analysis of the BPS equations are indeed the ones minimizing the localization action.

\subsubsection{Solving the BPS-equations}

To find the localization locus we study solutions of the BPS equations in their cohomological form of \eqref{eq:full_cohomological_susy_variations}. Before turning to the general case, we gain some intuition by considering the topological theory with $\epsilon_+^i = 0$. While we derived the complex in the presence of both $\epsilon_\pm^i$ as well as $\eta_-^i$, the equations include the $\eta_+^i = \eta_-^i = 0$ case. One simply sets\footnote{
  One could introduce an arbitrary $T^-$ since it is now a free parameter. We refrain from doing so.
}
\begin{equation}
  V = T^\pm = 0\,.
\end{equation}
Now, $F = \mathcal{F}$, and $\Psi$ depends only on $\Omega_+^i$, while $H = -2 \imath F^- + \frac{\imath}{2} \mathcal{M}_-^{ij} Y_{ij}$. The supersymmetry variations take the same form as in \eqref{eq:cohomological_SUSY_algebra}, except that the gauge parameter is now just $G_\phi$ and that $\delta \phi = 0$. Also, the Lie-derivatives vanish. So the complex is
\begin{equation}
  \delta Z = Z^\prime\,, \qquad
  \delta Z^\prime = G_\phi Z\,.
\end{equation}
Note that one can obtain the same complex by dimensional reduction of the Sasaki-Einstein complex \cite{Qiu:2013pta} along the Reeb vector. The scalars $\phi, \bar{\phi}$ are a linear combination of the five-dimensional real scalar $\sigma$ and the component of the five-dim.~gauge field along the Reeb.

Turning to the vanishing of the supersymmetry variations, the fermions $\eta, \Psi, \chi$ yield
\begin{equation}
  D\phi = [\phi, \bar{\phi}] = H = 0\,.
\end{equation}
Thus $4 F^- = \mathcal{M}_-^{ij} Y_{ij}$. Now the reality properties of $Y_{ij}$ are crucial. In \cite{Klare:2013dka}, they are $(Y_{ij})^* = Y^{ij}$. However, we rotate the countour of integration for the $SU(2)$-triplet by $90^\circ$ such that
\begin{equation}\label{eq:Yij_contour}
  (Y_{ij})^* = -Y^{ij}\,.
\end{equation}
This choice of contour also renders the $Y^i_{\phantom{i}j} Y^j_{\phantom{j}i}$ term in \eqref{eq:vector_multiplet_action} positive definite and thus convergent. Similar observations regarding contour choices and the convergence of the original path integral were made in \cite{Hama:2012bg,Kallen:2012va}. In order to further probe this choice, it is interesting to consider the topologically twisted theory. One can easily see that, with this choice, $F^-$ and $Y_{ij}$ decouple and
\begin{equation}
  Y_{ij} = 0, \qquad F^- = 0\,.
\end{equation}
We can now compare this saddle point configuration with the five-dimensional  $\mathcal{N}=1$ theories of \cite{Hosomichi:2012ek,Kallen:2012va,Qiu:2013pta}. Note that these references do consider an equivariant twist. However, the equivariant vector is the Reeb, along which one would naturally reduce to get the 4d topologically twisted theory.\footnote{Strictly speaking, \cite{Qiu:2013pta} allows for generic choices of Reeb while we assume for our argument that we are dealing with the canonical one.
}
In the 5d case, the theories generally localize to contact instantons, \textit{i.e.}~the gauge field satisfies equations like $\imath_R F = 0$ and $(1 - \imath_R \star) F = 0$. While it is in general not possible to simply reduce a generic contact instanton to an instanton and one has to be careful when comparing the two, it is still pleasing that the localization locus in the chiral theory takes essentially the same form, hence vindicating the contour \eqref{eq:Yij_contour}.

Returning to the full theory with $V \neq 0$ and $\epsilon_+^i \neq 0$, we again consider the vanishing of the supersymmetry variations. On the interior of the Delzant polytope, we know that $V \neq 0$, and we consider $\delta \eta = \delta \Psi = \delta\chi = 0$. These equations impose
\begin{equation}
  H = 0, \qquad
  D_V \bar{\phi} = [\phi, \bar{\phi}]\,, \qquad
  \imath_V \mathcal{F} + D\phi - V^2 D\bar{\phi} = 0\,.
\end{equation}
We study $H = 0$. As before, we consider the action of complex conjugation on
\begin{equation}
  \begin{aligned}
    \mathcal{F}^- = F^- - \frac{1}{2} \bar{\phi} T^- &= \frac{1}{4} \mathcal{M}_-^{ij} Y_{ij} - 2 (D\bar{\phi} \wedge V)^-\,.
  \end{aligned}
\end{equation}
With the reality condition for $Y_{ij}$ as in \eqref{eq:Yij_contour} we can decompose the real and imaginary parts as
\begin{equation}
\label{eq:Y}
  \begin{aligned}
    (F + \re \bar{\phi} dV)^- &= -2 (D \re\bar{\phi} \wedge V)^-\,, \\
    \imath \im \bar{\phi} dV^- &= \frac{1}{4} \mathcal{M}_-^{ij} Y_{ij} - 2 \imath (D\im\bar{\phi} \wedge V)^-\,.
  \end{aligned}
\end{equation}
Similarly we decompose the $\imath_V \mathcal{F}$ equation into
\begin{equation}
  \begin{aligned}
    0 &= \imath_V (F + \re \bar{\phi} dV) + D \re\phi - V^2 D \re\bar{\phi}\,, \\
    0 &= (\im \bar{\phi}) \imath_V dV + D \im \phi - V^2 D \im\bar{\phi}\,.
  \end{aligned}
\end{equation}
At this point one can compare the equations involving the gauge field to $(3.49)$ and $(3.50)$ in \cite{Gomis:2011pf}. In both cases, the reality conditions decouple the gauge field from the auxiliary modes, which again vindicates our contour choice \eqref{eq:Yij_contour}.

To proceed, we set
\begin{equation}
  \phi = \phi_1 + \imath \phi_2\,, \qquad
  \bar{\phi} = \phi_1 - \imath \phi_2\,, \qquad
  \phi_1, \phi_2 \in C^\infty(M_4)\,.
\end{equation}
The equation $D_V \bar{\phi} = [\phi, \bar{\phi}]$ then splits into real and imaginary parts
\begin{equation}
  D_V \phi_1 = 0\,, \qquad
  D_V \phi_2 = 2 [\phi_1, \phi_2]\,.
\end{equation}
In appendix \ref{sec:appendix_localization_locus} we adapt an argument from \cite{Gomis:2011pf} to show that the above equations for $F^-$ and $\imath_V F$ imply
\begin{equation}
  F + \phi_1 dV = 0\,.
\end{equation}
This is solved by
\begin{equation}
  \mathcal{A} = - \phi_1 V, \qquad D \phi_1 = d \phi_1 = 0\,.
\end{equation}

The other scalar $\phi_2$ satisfies two equations
\begin{equation}
  0 = (1 + V^2) D\phi_2 - \phi_2 \imath_V dV\,, \qquad
  D_V \phi_2 = [\phi_1, \phi_2]\,.
\end{equation}
It follows from the first of these that $D_V \phi_2 = 0$ so $[\phi_1, \phi_2] = 0$. Since $d(V^2) = - \imath_V dV$, the equation can be immediately integrated
\begin{equation}
  \phi_2 = \frac{\alpha_2}{1+V^2}, \qquad \alpha_2 \in \mathfrak{g}, \qquad [\alpha_2, \phi_1] = 0\,.
\end{equation}
To conclude, writing $\phi_1=\alpha_1$ with $\alpha_1\,\in\, \mathfrak{g}$, the relevant BPS configurations are
\begin{equation}\label{eq:BPS_solution}
  \phi_1 = \alpha_1\,, \qquad
  \phi_2 = \frac{\alpha_2}{1+V^2}\,, \qquad [\alpha_1,\,\alpha_2]=0\, , \qquad 
  \mathcal{A} = - \alpha_1 V.
\end{equation}
The value of the auxiliary triplet $Y_{ij}$ can then be directly read off from \eqref{eq:Y} and is given in appendix \ref{sec:appendix_tree_level}.

Consider now the gauge transformation $G_{\phi - V^2 \bar{\phi} - \imath_V \mathcal{A}}$ appearing in the supersymmetry algebra. Substituting the above solution leads to $G_{\alpha_1 + \imath \alpha_2}$. Hence, for the moment we are dealing with a complex gauge transformation. As we will see below, this changes once one considers the ghost sector as we will do in section \ref{sec:gauge_fixing}. Note as well that the gauge parameter, \textit{a priori} containing the $V^2$, becomes a constant once evaluated on the saddle configurations.

Note that the analysis we have so far performed is valid as long as $V^2\ne 0$. In turn, the points where $V^2=0$ must be studied separately. As one might suspect, new solutions will arise from those points. We will discuss them separately in section \ref{sec:instantons}.

\subsubsection{Localization action}
\label{sec:localizing_action}

One can recover the results from the previous section as the zero locus of the $\delta$-exact action
\begin{equation}
S_{\rm loc}= \delta \Big(\int {\rm Tr}(\,\bar{\Omega}_-^i\,\delta\Omega_-^i+\bar{\Omega}_+^i\,\delta\Omega_+^i\,)\,\Big)\, .
\end{equation}
Using the explicit form of the SUSY variations including the background Killing spinors the bosonic part can be written in a manifestly positive form as
\begin{equation}\label{eq:localization_action}
\bar{\delta\Omega_-^i}\,\delta\Omega_-^i+\bar{\delta\Omega_+^i}\,\delta\Omega_+^i= 2\,(1+V^{-2})\,|\imath_V\,D\phi|^2+2\, (1+V^2)\, [\phi,\,\overline{\phi}]^2 +\frac{1}{2}\,|M_-\,|^2+\frac{V^2}{2}\,|M_+\,|^2\, ,
\end{equation}
where
\begin{equation}
  \begin{aligned}
    M_+^{mn}&=(F^{mn})^+-\frac{\bar{\phi}}{2}\,(T^+)^{mn}+\frac{\imath}{2\,V^2}\,D_k\bar{\phi}\,\bar{\epsilon_+^i}\,\gamma^{mn}\,\gamma^k\,\epsilon_-^i+\frac{1}{4\,V^2}\,Y^i\,_j\,\bar{\epsilon_+^i}\,\gamma^{mn}\,\epsilon_+^j\, , \\
M_-^{mn}&=(F^{mn})^--\frac{\imath}{2}\,D_k\bar{\phi}\,\bar{\epsilon_-^i}\,\gamma^{mn}\,\gamma^k\,\epsilon_+^i-\frac{1}{4}\,Y^i\,_j\,\bar{\epsilon_-^i}\,\gamma^{mn}\,\epsilon_-^j +\frac{\imath}{4}\,\bar{\phi}\,\bar{\epsilon_-^i}\,\gamma^{mn}\,\slashed{\nabla}\epsilon_+^i\, .
  \end{aligned}
\end{equation}

In the $t\rightarrow\infty$ limit only the configurations for which $S_{\rm loc}=0$ contribute to the path integral. Upon separating the real and imaginary parts of the scalar field as $\phi=\phi_1+\imath\,\phi_2$, at a generic point, where $V^2\ne 0$, the zeros of $S_{\rm loc}$ are readily found as
\begin{eqnarray}
\begin{aligned}
0&=F_{mn}^--(D_{[m}\phi_1\,V_{n]})_- +\frac{\imath\,\phi_1}{4}\,\bar{\epsilon_-^i}\,\gamma^{mn}\,\slashed{D}\epsilon_+^i\, , \\ 
0&=  2\,\imath\,(D_{[m}\phi_2\,V_{n]})_--\frac{1}{2}\,Y^i\,_j\,\bar{\epsilon_-^i}\,\gamma^{mn}\,\epsilon_-^j +\frac{\phi_2}{2}\,\bar{\epsilon_-^i}\,\gamma^{mn}\,\slashed{D}\epsilon_+^i \, ,
\end{aligned}
\end{eqnarray}
from imposing $M_-=0$, and 
\begin{eqnarray}
\begin{aligned}
0&=F^+_{mn}-\frac{1}{V^2}\,(D_{[m}\phi_1\,V_{n]})_+-\frac{\phi_1}{2}\,T^+_{mn}\, ,  \\ 0&= 2\,\imath\,(D_{[m}\phi_2\,V_{n]})_+-\frac{1}{2}\,Y^i\,_j\,\bar{\epsilon_+^i}\,\gamma^{mn}\,\epsilon_+^j +\imath\,\phi_2\,V^2\,T^+\, ,
\end{aligned}
\end{eqnarray}
from $M_+=0$. Besides, we also have the conditions $[\phi,\,\bar{\phi}]=0$ and $\imath_VD\phi=0$. One can then verify that the solution to these equations is given by \eqref{eq:BPS_solution}. 

\subsubsection{Instanton configurations}
\label{sec:instantons}

By inspection of the localization action \eqref{eq:localization_action}, it is clear that, in addition to the configurations discussed above, we can have another whole family of solutions arising from the fixed points of the $U(1)$ action, where $V^2=0$, which must be studied separately. 

Considering the $V^2\,|M_+|^2$ term first, since the real part of $M_+$ contains a $V^{-2}\, D_{[m}\phi_1\,V_{n]}$, the localization action will contain a $V^{-2}\, D_{[m}\phi_1\,V_{n]}\,D_{[m}\phi_1\,V_{n]}$ term, which, at $V^2=0$ blows up unless we set $\phi_1=\alpha_1$ a Lie algebra-valued constant. Because of a similar argument, $\phi_2$ must also be set as well to a Lie algebra-valued constant $\phi_2=\alpha_2$, both subject to $[\alpha_1,\,\alpha_2]=0$. Furthermore, it is easy to convince oneself that the solution for the $Y$'s is $Y^1\,_2=Y^2\,_1=0$ and $Y^1\,_1=-Y^2\,_2=-\alpha_2\,|dV^-|$, where $dV^-$ is evaluated at the fixed points of the $U(1)$ action. In fact, one can check that, as for $\phi_1,\,\phi_2,\,Y^i\,_j$, these solutions are just the $V^2= 0$ limit of the generic $V^2\ne 0$ configurations. Finally, from the vanishing of $M_-$, we find an equation for the gauge field, which, using that $dV^-=J$ at the fixed points, can be re-written as $F^-+\alpha_1\,J=0$, with $J$ the Kahler form of $\CP^2$ evaluated at the fixed points. Note that, compared with the regular points for which $V^2\ne 0$, the ASD part of the gauge field equation drops out due to the $V^2$ factor multiplying $M_+$. Moreover, since the neighbourhoods of the fixed points of the $U(1)$ action are locally copies of $\mathbb{C}^2$, $J$ becomes the familiar constant K\"ahler form on flat space. It is then clear the equation $F^-+\alpha_1\,J_{\mathbb{C}^2}=0$ on $\mathbb{C}^2$ admits no finite energy solution unless $\alpha_1=0$. Hence, the relevant, finite energy, configurations around the points where $V^2=0$ are given by
\begin{equation}
\label{eq:instanton_loci}
F^-=0\, ,\quad \phi_1=0\, ,\quad \phi_2=\alpha_2\, ,\quad [\alpha_1,\,\alpha_2]=0\, ,\quad Y^i\,_i=-\frac{1}{2}\,\alpha_2\,|dV^-|\, ,\quad Y^1\,_2=Y^2\,_1=0\, .
\end{equation}
Note that, in our conventions, $F^-=0$ implies $F_{mn}=-\frac{1}{2}\,\epsilon_{mnab}\,F^{ab}$, while the $J$ on $\CP^2$ satisfies $J_{mn}=\frac{1}{2}\,\epsilon_{mnab}\,J^{ab}$. Hence the $V^2=0$ points support localized anti-instanton solutions.

Note as well that the above configuration seems, at first sight, a bit at odds with that at generic points, as the latter seems to involve a non-zero $\alpha_1$ while the former demands a vanishing $\alpha_1$. As we will see in the next subsection, this is resolved once the ghost sector is taken into account.

\subsection{Gauge fixing}\label{sec:gauge_fixing}

The BRST complex and gauge fixing work in the same way as in \cite{Pestun:2007rz,Hama:2012bg,Kallen:2011ny,Kallen:2012cs}. For early accounts of ghosts for ghosts in gauge theories, see \cite{Blau:1989bq,Baulieu:1996rp} and references therein. Carrying things over to our conventions, we define (see eq.\eqref{eq:cohomological_complex})
\begin{equation}
  Z = (\bar{\phi}, \mathcal{A}, \chi)\, , \qquad
  Z^\prime = (\eta, \Psi, H)\,.
\end{equation}
and include a ghost sector $(c, \tilde{c}, b, c_0, \tilde{c}_0, a_0, \tilde{a}_0, b_0)$. Here $c$ and $\tilde{c}$ are ghost and anti-ghost (both fermionic), $b$ is a Lagrange multiplier (bosonic), all remaining fields are introduced to deal with the zero modes. Out of these, $c_0$ and $\tilde{c}_0$ are fermionic, the rest bosonic. For convenience, we define
\begin{equation}
  \sigma \equiv \phi - V^2 \bar{\phi} - \imath_V \mathcal{A} = (1-V^2) \phi_1 - \imath_V \mathcal{A} + \imath (1+V^2) \phi_2\,.
\end{equation}
The supersymmetry variations of the full system are
\begin{equation}
  \begin{aligned}
    \delta_S c &= -\sigma\,, &
    \delta_S \tilde{c} &= 0\,, &
    \delta_S c_0 &= 0\,, &
    \delta_S \tilde{c}_0 &= 0\,, \\
    \delta_S a_0 &= 0\,, &
    \delta_S \tilde{a}_0 &= 0, &
    \delta_S b &= \pounds_V \tilde{c}\,, &
    \delta_S b_0 &= 0\,, \\
    & &
    \delta_S \sigma &= 0\,, &
    \delta_S Z &= Z^\prime\,, &
    \delta_S Z^\prime &= \pounds_V Z + G_{\sigma} [Z]\,.
  \end{aligned}
\end{equation}
In addition, we define the BRST transformations
\begin{equation}
  \begin{aligned}
    \delta_B c &= a_0 - \frac{g}{2} [c, c]\,, &
    \delta_B \tilde{c} &= b\,, &
    \delta_B c_0 &= G_{a_0} b_0\,, &
    \delta_B \tilde{c}_0 &= G_{a_0} \tilde{a}_0\,, \\
    \delta_B a_0 &= 0\,, &
    \delta_B \tilde{a}_0 &= \tilde{c}_0\,, &
    \delta_B b &= G_{a_0} \tilde{c}\,, &
    \delta_B b_0 &= c_0, \\
    & &
    \delta_B \sigma &= - \pounds_V c - g [c, \sigma]\,, &
    \delta_B Z &= G_c Z\,, &
    \delta_B Z^\prime &= G_c Z^\prime\,.
  \end{aligned}
\end{equation}
Then, upon forming
\begin{equation}
  \rho = a_0 - \sigma - \frac{g}{2}[c,c]\,, \qquad
  S = Z' + G_c Z\, ,
\end{equation}
and considering the ``vectors''
\begin{equation}\label{eq:cohomological_complex_including_ghosts_multiplets}
  Y = (Z, c, \tilde{c}, b_0, \tilde{a}_0)\,, \qquad
  Y' = (S, \rho, b, c_0, \tilde{c}_0)\,,
\end{equation}
we find
for $\delta = \delta_S + \delta_B$
\begin{equation}\label{eq:cohomological_complex_including_ghosts}
    \delta Y =  Y'\,, \qquad
    \delta Y' = (\pounds_v+ G_{a_0}) Y\,, \qquad
    \delta a_0 = 0\,.
\end{equation}

To fix the gauge we add the term $\delta V_{\text{g.f.}}$ to the action. $V_{\text{g.f.}}$ is essentially the same as in \cite{Pestun:2007rz}, yet with $\xi_1 = 0$. See also \cite{Kallen:2011ny}. In detail (with $\xi_2 > 0$)
\begin{equation}\label{eq:gauge-fixing_term}
  V_{\text{g.f.}} = \left(\tilde{c}, \imath d^\dagger \mathcal{A} + \imath b_0 \right) + \left(c, \tilde{a}_0 - \frac{\xi_2}{2} a_0 \right)\,.
\end{equation}
Then,
\begin{equation}
  \begin{aligned}
    \delta V_{\text{g.f.}} &= \imath (b, d^\dagger \mathcal{A}) - \imath (\tilde{c}, d^\dagger \Psi) - \imath (\tilde{c}, d^\dagger d_{\mathcal{A}}c) \\
    &+ \imath (b, b_0) - \imath (\tilde{c}, c_0) - (c, \tilde{c}_0) + \left(\rho,  \tilde{a}_0 - \frac{\xi_2}{2} a_0\right)\,.
  \end{aligned}
\end{equation}
We need to verify that this is positive definite, and consider the terms involving $a_0$:
\begin{equation}
  \left(\rho, \tilde{a}_0 - \frac{\xi_2}{2} a_0\right) = - \frac{\xi_2}{2} \left(a_0 - \sigma - \frac{g}{2} [c,c], a_0 - \frac{2}{\xi_2} \tilde{a}_0 \right)\,.
\end{equation}
Wick rotating $a_0$, we set $a_0 = \imath a_0^E$ with $a_0^E \in \bR$. Performing the integral over $a_0^E$,
\begin{equation}
  \frac{\xi_2}{2} \left(a_0^E + \imath \sigma + \frac{\imath g}{2} [c,c], a_0^E + \frac{2\imath}{\xi_2} \tilde{a}_0 \right)
  \to \frac{1}{2\xi_2} \left[ \tilde{a}_0 - \frac{\xi_2}{2} \left( \sigma + \frac{g}{2} [c,c] \right) \right]^2.
\end{equation}
The partition function is independent of $\xi_2$. At $\xi_2 = 0$,
\begin{equation}
  \left( \imath a^E_0 - \sigma - \frac{g}{2} [c,c], \tilde{a}_0 \right)
\end{equation}
we integrate $\tilde{a}_0$ out we find that
\begin{equation}
  a_0^E = \im \sigma = (1+V^2) \phi_2 = \alpha_2\,, \qquad 0 = \re \sigma = (1-V^2) - \imath_V \mathcal{A}= \alpha_1\,.
\end{equation}
The other terms in $\delta V_{\text{g.f.}}$ are dealt with as in \cite{Pestun:2007rz}.

Regarding the localization locus, consider
\begin{equation}
  \delta c = a_0 - \imath (1+\imath V^2) \phi_2 - (1-V^2) \phi_1 + \imath_V \mathcal{A} - \frac{g}{2} [c, c]\,.
\end{equation}
Per usual, the previous results on the localization locus (or the BPS solutions) are unaffected. Thus we substitute \eqref{eq:BPS_solution} and obtain
\begin{equation}
  \delta c = a_0 - (\alpha_1 + \imath \alpha_2) - \frac{g}{2} [c, c]\,.
\end{equation}
For the right hand side to vanish, we need $a_0 = \alpha_1 + \imath \alpha_2$. Depending on the reality condition for $a_0$, one of the two constant factors is set to zero. Choosing the contour such that $a_0 = \imath a_0^E$, we obtain
\begin{equation}\label{eq:ghost_sector_localization_locus}
  \alpha_1 = 0, \qquad a_0^E = \alpha_2\,.
\end{equation}
Note that this has the additional effect of setting to zero the background gauge field in the localization locus at generic points \eqref{eq:BPS_solution}, in parallel with the instanton solutions in \eqref{eq:instanton_loci}. In addition, the gauge transformation parameter $\phi-V^2\bar{\phi}-\imath_V\,\mathcal{A}$ becomes, as expected, purely imaginary (and subsequently purely real upon the Wick rotation) and constant. Moreover, this nicely reconciles with the instanton sector, which demanded $\alpha_1=0$ to find finite action configurations. Note that these saddle points correspond to configurations with vanishing first Chern class -- i.e.~$F=0$.

The action \eqref{eq:gauge-fixing_term} is not unique. Changing the sign of the second term, one finds that it is necessary to Wick rotate $\tilde{a}_0$ instead of $a_0$. In this case it follows that $\alpha_2 = 0$ while $\alpha_1 = a_0$. Hence, from \eqref{eq:BPS_solution} it follows that there is a background field $\mathcal{A}= - \alpha_1\, V$. A priori there seems to be nothing that keeps us from making this choice. By explicit computation one finds that our results for the perturbative partition function would be different. The instanton sector would exhibit as well crucial differences. Recall that, in order to have finite energy configurations coming from the $V^2=0$ loci we needed to demand $\alpha_1=0$. Hence the instanton sector would only contribute upon choosing  \eqref{eq:gauge-fixing_term}. We will come back to this issue below.

\section{The partition function}\label{sec:partition_function}

As outlined above, upon taking the classical limit in $t$, the spurious $\hbar$ introduced by the localization action, the partition function can be exactly computed by saddle point approximation. Hence, it acquires contributions only from the localization locus; each being the product of the classical action evaluated at the locus times the fluctuation determinant. Since there are two types of loci, namely the perturbative configurations arising from $V^2\ne 0$ and the instanton configurations sitting at $V^2=0$, the partition function takes the form

\begin{equation}
\label{eq:Z}
  \int_{\mathfrak{g}} [da_0^E] \, Z_{\text{cl}} \, Z_{\text{1-loop}} \, Z_{\text{instantons}}\,.
\end{equation}

In order to compute the various ingredients, we follow \cite{Qiu:2013pta,Schmude:2014lfa,Kallen:2011ny,Kallen:2012cs}. Actually, the situation is slightly simpler than in \cite{Pestun:2007rz,Hama:2012bg} since we do not have to worry about an operator $D_{10}$ vanishing on the horizon. As in \cite{Kallen:2011ny}, we make use of the Weyl integration formula (see e.g.~\cite{goodman2009symmetry}). Then

\begin{equation}
\label{eq:ZWeyl}
  \frac{1}{\vert W \vert} \frac{\vol G}{\vol T} \int_{\mathfrak{t}} [da_0^E] \, \prod_{\beta > 0} \langle a_0^E, \beta \rangle^2\,  Z_{\text{cl}}(a_0^E)\,  Z_{\text{1-loop}}(a_0^E) \, Z_{\text{instantons}}(a_0^E)\,.
\end{equation}
Note that a side effect of \eqref{eq:ghost_sector_localization_locus} is that the integral in \eqref{eq:Z} or \eqref{eq:ZWeyl}, which otherwise would have been over the whole complex plane spanned by $\alpha_1+\imath\,\alpha_2$, gets projected to the real line.

In the following we will discuss each of the terms in \eqref{eq:Z} individually.

\subsection{Tree level contribution}\label{sec:tree_level_contribution}

For our background the action \eqref{eq:vector_multiplet_action} reduces to
\begin{equation}
  \mathcal{L} = \frac{1}{8} Y^i_{\phantom{i}j} Y^j_{\phantom{j}i} + (d\phi_1)^2 + (d\phi_2)^2 - g [\phi, \bar{\phi}] + \frac{1}{8} F_{mn} F^{mn} 
  - \frac{1}{4} \phi F_{mn} T^{+mn}
  - \frac{1}{16} \phi^2 T^+_{mn} T^{+mn}\, .
\end{equation}
Here one should note that while we redefined the gaugino variations such that there are both $T^+$ and $T^-$, this redefinition does not affect the action \eqref{eq:vector_multiplet_action}. Hence we need to use $T^- = 0$ when studying the above action. Evaluating this at the localization locus given by \eqref{eq:BPS_solution} and \eqref{eq:ghost_sector_localization_locus} one finds,
\begin{equation}\label{eq:classical_action_in_terms_of_v__dv_over_v-squared}
  S_{\text{cl}} = \frac{(a_0^E)^2}{4 g_{\text{YM}}^2} \int_{M_4} \vol \left( \frac{dV}{1+V^2}\right)^2\,, \qquad
  Z_{\text{cl}}(a_0^E) = e^{- S_{\text{cl}}}\, ,
\end{equation}
as we show in appendix \ref{sec:appendix_tree_level}. 

It appears as if \eqref{eq:classical_action_in_terms_of_v__dv_over_v-squared} might depend on the metric. However, since our derivation assumed from the start that the manifold $M$ is toric K\"ahler, the metric is directly related to the complex structure. By direct calculation one can establish the dependence on the potential $g(x)$ appearing in the construction of Guillemin and Abreu \cite{guillemin1994kaehler,abreu2003kahler}, yet this corresponds to different choices of K\"ahler potential. The situation appears to be similar to that when comparing the partition functions on the four-sphere \cite{Pestun:2007rz} and the ellipsoid \cite{Hama:2012bg}, where the overall result shows a clear dependence on the squashing parameters. Note as well that the $V^2$ dependence was already a feature of the supersymmetry complex while the appearance of the $1 + V^2$ term can also be thought of in terms of the norms of both spinors $\epsilon^i_\pm$.

Evaluating \eqref{eq:classical_action_in_terms_of_v__dv_over_v-squared} for $\CP^2$ using the canonical metric and symplectic structure given by the potential $g(x) = g_P(x)$, we find
\begin{equation}\label{eq:Zcp2_tree-level_integrand}
  \begin{aligned}
    \left( \frac{dV}{1+V^2} \right)^2 &= 8 \frac{p^2 (5 x_1^2 - 4 x_1 + 1) + q^2 (5 x_2^2 - 4 x_2 + 1) + 2pq (5 x_1 x_2 - x_1 - x_2)}{[2p^2 (x_1^2 - x_1) + 2 q^2 (x_2^2 - x_2) + 4 pq x_1 x_2 - 1]^2} \equiv I_{\CP^2}\,.
  \end{aligned}
\end{equation}
Thus we can calculate the integral using the measure
\begin{equation}
  \begin{aligned}
    \int_{\CP^2} \vol = \int_0^1 dx_2 \int_0^{1-x_2} dx_1 \int_0^{2\pi} dy_1 \int_0^{2\pi} dy_2\,.
  \end{aligned}
\end{equation}
In the end, the overall result is
\begin{equation}\label{eq:Zcp2_tree-level_contribution}
  \begin{aligned}
  S_{\text{cl}}^{\CP^2} = 
    &\frac{4 (a_0^E)^2 \pi^2}{g_{\text{YM}}^2} \frac{1}{pq (p-q) \sqrt{(p^2+2)(q^2+2)[(p-q)^2+2]}} \\
    \Bigg\{ &- \sqrt{(q^2 + 2)[(p-q)^2 + 2]} (5 p^2 - 2 pq + 2q^2 + 9) \arctanh \frac{p}{\sqrt{p^2 + 2}} \\
    &+ \sqrt{(p^2 + 2)[(p-q)^2 + 2]} (5q^2 - 2pq + 2p^2 + 9) \arctanh \frac{q}{\sqrt{q^2+2}} \\
    &+ \sqrt{(p^2+2)(q^2+2)} (5p^2 - 8pq + 5q^2 + 9) \arctanh \frac{p-q}{\sqrt{(p-q)^2 + 2}} \Bigg\}\,.
  \end{aligned}
\end{equation}

\subsection{One-loop contribution}\label{sec:one-loop_contribution}

Following the localization argument, the fluctuation determinant is
\begin{equation}
  Z_{\text{1-loop}}(a_0^E) = \frac{\det_{\text{fermions}} \delta^2}{\det_{\text{bosons}} \delta^2}\,,
\end{equation}
with $\delta^2$ given by \eqref{eq:cohomological_complex_including_ghosts}. In opposite to \cite{Pestun:2007rz,Hama:2012bg}, we can evaluate the above directly just as in \cite{Schmude:2014lfa}. The fermions appearing in $Y$ are $\chi \in \Omega^- = \Omega^{2,0} \oplus \Omega^{0,0} J \oplus \Omega^{0,2}$, as well as $c, \bar{c} \in \Omega^{0,0}$. The bosonic modes are $\phi_1 \in \Omega^{0,0}$, $W \in \Omega^{1,0} \oplus \Omega^{0,1}$ as well as the zero-modes $b_0, \bar{a}_0$, which are harmonic functions. Of course we mean $\Omega^{p,q} = \Omega^{p,q}(M, \mathfrak{g})$. $\phi$ is not included here as it is not among the ``coordinates'' \eqref{eq:cohomological_complex_including_ghosts_multiplets}. We will deal with it in the final matrix integral. Thus we want to calculate
\begin{equation}
  Z_{\text{1-loop}}(a_0^E) = \sqrt{ \frac{\det_L \Omega^{2,0} \det_L \Omega^{0,0}}{\det_L \Omega^{1,0}} }
  \sqrt{ \frac{\det_L \Omega^{0,2} \det_L \Omega^{0,0}}{\det_L \Omega^{0,1}} }
  \frac{1}{\det_L H^0}\,,
\end{equation}
where $L = \pounds_V + \imath G_{a_0^E}$ and we have changed the notation $\det_A B \to \det_B A$ for readability.

There are no non-trivial harmonic forms on a compact K\"ahler manifold, so we drop the last term. Then the evaluation of the above is based on the fact that we have effectively two copies of the Dolbeault complex
\begin{equation}
  \dots \xrightarrow{\bar{\partial}} \Omega^{0,q-1} \xrightarrow{\bar{\partial}} \Omega^{0,q} \xrightarrow{\bar{\partial}} \Omega^{0,q+1} \xrightarrow{\bar{\partial}} \dots\,.
\end{equation}
Now, any form $\eta \in \Omega^{0,q-1}$ defines a form $\bar{\partial} \eta \in \Omega^{0,q}$. These cancel in the alternating product unless $\eta$ is holomorphic. Next one has only to consider elements $\psi \in \Omega^{0,q}$ that don't descent from $\Omega^{0,q-1}$; i.e.~that are not exact. Again they cancel against their descendants $\bar{\partial} \psi \in \Omega^{0,q+1}$ unless they are holomorphic. So we are counting holomorphic modulo exact forms and the result is the alternating quotient
\begin{equation}\label{eq:1-loop_partition_function_and_cohomology}
  Z_{\text{1-loop}}(a_0^E) = \sqrt{\frac{\det_L H^{0,2} \det_L H^{0,0}}{\det_L H^{0,1}}}
  \sqrt{\frac{\det_L H^{2,0} \det_L H^{0,0}}{\det_L H^{1,0}}}\,.
\end{equation}
Once again we note that this is formally identical to the Sasaki-Einstein case with Dolbeault cohomology taking the role of Kohn-Rossi cohomology \cite{Schmude:2014lfa}. Now, we know that $h^{0,0} = 0$. Moreover, $h^{1,0} = \frac{1}{2} b_1 = 0$ and $h^{2,0} = 0$ unless $M$ is Calabi-Yau. Thus we conclude that
\begin{equation}
  Z_{\text{1-loop}}(a_0^E) = 1\,.
\end{equation}
This agrees with \cite{Bawane:2014uka} in the special case $m = n = 0$. Furthermore, it is also consistent with the 5d result in \cite{Kallen:2012cs}, which becomes non-trivial only when the $S^1$ is fibered on top of the $\CP^2$ so as to make an $S^5$. See also \cite{Kim:2012qf}.

Again one can expect the results of this section to change when choosing the alternate ghost contour $\tilde{a}_0 = \imath \tilde{a}_0^E$. Due to the background field $F = - \alpha_1 dV$ we could for example no longer link the zero modes of $d^\dagger d_{\mathcal{A}}$ to harmonic functions.

\subsection{Instantons}
\label{sec:instanton_contribution_to_action}

In addition to the $V^2\ne 0$ saddle points of the localization action we have extra saddle points sitting at the loci where $V^2=0$. As we have discussed, we are considering a generic $V$ such that the set $V^2=0$ contains a discrete and isolated number of points, around which the space looks like a copy of $\mathbb{C}^2$. As discussed in section (\ref{sec:localizing_action}), the relevant configurations sitting at $V^2=0$ are given by the equation $F^-+\alpha_1\,J_{\mathbb{C}^2}=0$, with $J_{\mathbb{C}^2}$ the K\"ahler form on $\mathbb{C}^2$. However, the solution to this equation on $\mathbb{C}^2$ does not yield finite energy (action) configurations unless $\alpha_1=0$, in which case the equation becomes the familiar $F_{mn}=-\frac{1}{2}\,\epsilon_{mnab}\,F^{ab}$. Hence we have anti-instanton configurations only contributing upon setting $\alpha_1=0$. This fits nicely with our choice of gauge-fixing action \eqref{eq:gauge-fixing_term} which restricts the perturbative solutions to the subset $\alpha_1=0$. Recall that our $V^2=0$ configurations are just the $V=0$ limit of those in \eqref{eq:BPS_solution} (dropping of course the ASD part in the gauge field equation). Hence the gauge-fixing choice not only projects the gauge parameter to be purely imaginary (as otherwise it would have been $\alpha_1+\imath\,\alpha_2$) but it is also such that it allows for anti-instantons located at $V^2=0$.

The configurations sitting at the points $V^2=0$ are given by eq.~\eqref{eq:instanton_loci}. By inspection, one can convince oneself that, in the neighbourhood of any point $V^2=0$, the background becomes a copy of the $\Omega$ background \cite{Nekrasov:2002qd,Nekrasov:2003rj} with equivariant parameters given, at each of them, by (see appendix \ref{sec:appendix_abreu})

\begin{equation}
\label{eq:epsilons}
  \begin{aligned}
    (\epsilon^{(1)}_1, \epsilon^{(1)}_2) &= (p,\, q) & &\qquad & (x^1, x^2) &= (0,0), \\
    (\epsilon^{(2)}_1, \epsilon^{(2)}_2) &=(q-p,\,-p) & &\qquad & (x^1, x^2) &= (1,0), \\
    (\epsilon^{(3)}_1, \epsilon^{(3)}_2) &=(-q,\,p-q) & &\qquad & (x^1, x^2) &= (0,1).
  \end{aligned}
\end{equation}
Since each fixed point is a copy of the $\Omega$ background, the contribution of each is a copy of the Nekrasov instanton partition function $Z^{\rm Nekrasov}(\epsilon_1,\,\epsilon_2,\,a_0^E)$. Explicit expressions for $Z^{\rm Nekrasov}$ have been computed in the literature for all the classical groups (see \textit{e.g.}~\cite{Shadchin:2005mx} for a thorough introduction and compilation of results). Hence

\begin{equation}
Z_{\rm instantons}(a_0^E)=\prod_{i=1}^3 Z^{\rm Nekrasov}(\epsilon_1^{(i)},\,\epsilon_2^{(i)},\,a_0^E)\, .
\end{equation}

The apparent factorization extends to the classical part as well. Upon inspection of the classical action in \eqref{eq:Zcp2_tree-level_contribution}, we observe that it can be neatly re-written as

\begin{equation}
S_{\rm cl}^{\CP^2}=\sum_{i=1}^3\,S^0(\epsilon_1^{(i)},\,\epsilon_2^{(i)},\,a_0^E)\, ;
\end{equation}
where the function $S^0(\epsilon_1,\,\epsilon_2,\,a_0^E)$ is given by

\begin{equation}
S^0(\epsilon_1,\,\epsilon_2,\,a_0^E)=\frac{4 (a_0^E)^2 \pi^2}{g_{\text{YM}}^2}\,\frac{9+5\,\epsilon_1^2-8\,\epsilon_1\,\epsilon_2+5\,\epsilon_2^2}{\epsilon_1\,\epsilon_2\,(\epsilon_1-\epsilon_2)\,\sqrt{(\epsilon_1-\epsilon_2)^2+2}}\,\arctanh\Big(\frac{\epsilon_1-\epsilon_2}{\sqrt{2+\,(\epsilon_1-\epsilon_2)^2}}\Big)\, .
\end{equation}
Therefore the classical contribution to the partition function splits into three contributions as $Z_{\rm cl}=\prod_{i=1}^3\,Z^0_{\rm cl}(\epsilon_1^{(i)},\,\epsilon_2^{(i)},\,a_0^E)$, with $Z_{\rm cl}^0(\epsilon_1,\,\epsilon_2,\,a_0^E)=e^{-S^0(\epsilon_1,\,\epsilon_2,\,a_0^E)}$. Keeping in mind that $Z_{\rm 1-loop}=1$, we see that the whole integrand in the partition function takes a factorized form

\begin{equation}
\label{eq:Zfactorized}
Z_{\rm cl}\,Z_{\rm 1-loop}\,Z_{\rm instantons}=\prod_{i=1}^{\rm corners} \, Z^0_{\rm cl}(\epsilon^{(i)}_1,\,\epsilon_2^{(i)},\,a_0^E)\,\,Z^{\rm Nekrasov}(\epsilon_1^{(i)},\,\epsilon_2^{(i)},\,a_0^E).
\end{equation}
The product runs to the solutions of $V^2=0$, that is, the corners of the Delzant polytope where, for the case of $\CP^2$, the corresponding values of the $\epsilon_{1,\,2}^{(i)}$ are given by \eqref{eq:epsilons}. Even though the instanton part of each of the corner contributions is just that of flat space, the classical and one-loop contributions do not coincide with their $\mathbb{C}^2$ counterparts. While the one-loop piece, which in our case is just 1, might admit more suited factorizations, the classical contribution, involving $\arctanh$, seems harder to understand along the lines of factorization in 3d and 5d (see \textit{e.g.} \cite{Pasquetti:2011fj,Nieri:2013yra,Nieri:2013vba,Qiu:2013aga,Qiu:2014oqa} for recent developments in a similar context). We leave this problem open for future research.

\section{Conclusions}\label{conclusions}

In this paper we have studied pure $\mathcal{N}=2$ supersymmetric gauge theories on toric K\"ahler manifolds, concentrating in particular on the case of $\CP^2$. By equivariantly twisting with a Killing vector generating part of the $\mathbb{T}^2$ action, we computed the supersymmetric partition function of the theory using localization. As one might expect, we found the theory localizes to a purely perturbative path integral in the bulk augmented by instantons situated at the fixed-points of the $\mathbb{T}^2$ action. While the explicit $V$ dependence appearing in the localization equations \eqref{eq:BPS_solution} might seem puzzling, we noted that it simplifies the gauge transformation appearing in the supersymmetry complex \eqref{eq:cohomological_complex} to a complex gauge transformation.

The partition function has contributions from the classical action, one-loop determinant and instanton sector. We found the one-loop determinant to be trivial as fermionic and bosonic modes cancel exactly. This is in agreement with both \cite{Bawane:2014uka} and the dimensional reduction of five-dimensional results. In the latter case, the one-loop function counts functions that are holomorphic with respect to the so-called tangential Cauchy-Riemann operator that are charged along the Reeb \cite{Schmude:2014lfa}. When reducing along the Reeb, all the charged modes should be discarded and one finds agreement with our result. Returning to the four-dimensional case, we showed that the classical contribution can be factorized into contributions arising from the $V^2=0$ loci. Hence, we find a natural factorization of the partition function which remains to be fully understood. Note that, assuming factorization for the one-loop action as well, we could consider the index of the self-dual complex \cite{Gomis:2011pf} at the $V^2=0$ points. It is easy to see that this sum is just a constant, in agreement with the trivial one-loop determinant which we find. In fact this might provide a better understanding of \eqref{eq:Zfactorized}. It should be remarked however that the factorization observed here for the classical action can be thought of as a direct consequence of our use of the Delzant construction. As we followed \cite{abreu2003kahler}, the complex structure and metric receive contributions from each vertex in the polytope. Thus it is to be expected that the overall result respects this structure.

The form of the classical action is the most surprising result. It should be noted however that all the ingredients appearing in \eqref{eq:classical_action_in_terms_of_v__dv_over_v-squared} were somewhat to be expected. The combination $1 + V^2$ is the sum of the norms of both spinors while $dV$ appeared naturally after eliminating the spinors $\eta^i_\pm$.

A crucial question is of course whether \eqref{eq:classical_action_in_terms_of_v__dv_over_v-squared} depends on the metric. Naively it might appear so. Yet, as we have described, this dependence is somewhat spurious, as we are dealing with a rather constrained space as it is a toric K\"ahler manifold and the metric is equivalent to the complex structure. Nevertheless it would be important to fully clarify this point.

Since our results follow directly from supergravity via rigid supersymmetry, it is important to point out that our result for the classical action appears different from that of \cite{Bawane:2014uka} whose authors did not take the supergravity approach. Note that the evaluation of the classical action is also intimately related to factorization, hence providing yet another motivation for a further study of this point.

A technical point ubiquitously appearing throughout this work is the question of choosing appropriate contours. We encountered this both when considering the contour of the auxiliary triplet $Y_{ij}$ and that of the ghosts. While our choice for the former -- \eqref{eq:Yij_contour} -- is in agreement with expectations from \cite{Gomis:2011pf,Hama:2012bg,Kallen:2012va,Qiu:2013pta}, it is not the only choice that renders the action \eqref{eq:vector_multiplet_action} positive definite. Indeed, one could in principle rotate the contour by up to $45^\circ$ from \eqref{eq:Yij_contour}. This would allow for mixing between $F$ and $Y_{ij} \mathcal{M}^{ij}_-$ and thus with the K\"ahler form.

The last point is especially interesting as it raises the question of how to generalize our results to non-vanishing first Chern class in order to make a better connection with \cite{Gottsche:2006tn,Bawane:2014uka}. Alternatively one could wonder whether it is appropriate to add an operator $e^{\int F \wedge \omega}$ to the path integral, where $\omega$ is a suitably chosen two-form.

A natural extension would be to incorporate matter hypermultiplets. It would be very interesting to check wether their contribution to the one-loop determinant is $1$ as well, as one might be tempted to conclude, if factorization is assumed, by summing the index of the Dirac complex \cite{Gomis:2011pf} at the corners of the $\CP^2$ Delzant polytope. Comparison with the Sasakian case does once again suggest so \cite{Qiu:2013aga,Qiu:2014oqa}.

In \cite{Gerchkovitz:2014gta,Gomis:2014woa} it was argued that the $S^4$ $\mathcal{N}=2$ partition function for SCFT's computes the K\"ahler potential on the conformal manifold. Likewise, it would be very interesting to elucidate the physical meaning of the $\CP^2$ (or generic toric K\"ahler manifold) partition function -- possibly in the conformal case, upon the addition of the suitable matter content.

We have assumed the Killing vector $V$ to be generic, so that $V^2=0$ only happens at a discrete set of points corresponding to the corners of the Delzant polytope. Nevertheless, for particular choices of $V$, we can have more general situations where $V^2$ vanishes over a whole $\CP^1$ corresponding to an edge of the polytope. In the case of $\CP^2$ one such case is $p_{\phi}=0$ or $p_{\psi}=0$ in eq.\eqref{eq:localV2}, which corresponds to $|\epsilon_1|=|\epsilon_2|$. Hence, the familiar case of $\epsilon_1=\epsilon_2$ belongs to the cases which, strictly speaking, are excluded from our computation. It would be interesting to understand this point better.

\section*{Acknowledgements}

The authors would like to thank Stefano Cremonesi, Jaume Gomis, Alexander Sasha Getmanenko, Amihay Hanany, Simeon Hellerman, Dario Martelli, Mauricio Romo, Eric Sharpe, Yu\-ji Ta\-chi\-ka\-wa, Alessandro Tomassiello, Alberto Zaffaroni and especially Patrick Meessen who contributed during the initial stages of this project. J.S.~is thankful for warm hospitality at Oxford, at CERN in the context of the CERN-Korea Theory Collaboration
funded by National Research Foundation (Korea), at Kavli IPMU, and at Perimeter Institute during various stages of this project. The authors are partly supported by the spanish grant MINECO-13-FPA2012-35043-C02-02. In addition, they acknowledge financial support from the Ramon y Cajal grant RYC-2011-07593 as well as the EU CIG grant UE-14-GT5LD2013-618459. They would also like to acknowledge support from the COST action COST-MP1210. J.S.~is also supported by the Asturias government Clar\'in FICYT grant ADC14-27.

\appendix

\section{Conventions}\label{sec:conventions}

\subsection*{General conventions}

Our conventions are to a large extend as in \cite{vanProyen,Freedman:2012zz,Klare:2013dka}. Self-duality and the Hodge dual are defined as (note the normalisation of $\Omega^\pm$):
\begin{equation}\label{eq:Hodge-star}
  \star \Omega_{ab} = -\frac{1}{2} \epsilon_{abcd} \Omega^{cd}\,, \qquad
  \Omega^\pm = \frac{1}{2} (\Omega \pm \star \Omega)\,.
\end{equation}
Then,
\begin{equation}
   \epsilon_{ab}^{\phantom{ab}cd} \Omega^\pm_{cd} = \mp 2 \Omega_{ab}^\pm\,.
\end{equation}
It's worthwhile to point out that due to the sign in $\star$, we have
\begin{equation}\label{eq:hodge_inner_product_sign}
  \forall \alpha, \beta \in \Omega^2, \qquad
  \alpha \wedge \star \beta = \star\beta \wedge \alpha = - \frac{1}{2} \alpha_{ab} \beta^{ab} \vol_4\,,
\end{equation}
so $\alpha \wedge \star \alpha$ is negative definite.

Turning to Yang-Mills theory, field strengths, covariant derivatives and transformations are
\begin{equation}\label{eq:Yang-Mills_conventions}
  \begin{aligned}
    F &= d\mathcal{A} + \mathcal{A} \wedge \mathcal{A}\,, &\qquad
    D_X\phi &= \nabla_X\phi + g [\mathcal{A}_X, \phi]\,, \\
    \delta_\theta \mathcal{A}_X &= \nabla_X \theta + g [\mathcal{A}_X, \theta] \equiv G_\theta [\phi]\,, &\qquad
    \delta_\theta \phi &= - g [\theta, \phi] \equiv G_\theta [\phi]\,.
  \end{aligned}
\end{equation}

\subsection*{Spinors}

For explicit calculations, we choose a basis
\begin{equation}
  \begin{aligned}
    \gamma_1 = \sigma_1 \otimes 1\,, \quad
    \gamma_2 = \sigma_2 \otimes 1\, ,\quad
    \gamma_3 = \sigma_3 \otimes \sigma_1\, ,\quad
    \gamma_4 = \sigma_3 \otimes \sigma_2\, .
  \end{aligned}
\end{equation}
Then, $\gamma_m^* = \gamma_m^T$ and $\gamma = \gamma^{1234} = -\sigma_3 \otimes \sigma_3$. Some useful standard identities involving chiral spinors $\chi_\pm$ are
\begin{equation}
  \begin{aligned}
    \gamma^n \chi_\pm &= \mp \frac{1}{3!} \epsilon^{klmn} \gamma_{klm} \chi_\pm\, , &\quad
    \gamma_{klm} \chi_\pm &= \mp \epsilon_{klmn} \gamma^n \chi_\pm\,, &\quad
    \gamma^{mn} \chi_\pm &= \mp \frac{1}{2} \epsilon^{klmn} \gamma_{kl} \chi_\pm\,.
  \end{aligned} 
\end{equation}
The conjugation matrix is given by $B = \imath \gamma^{24}$ and signs work out such that $B \gamma_m B^{-1} = \gamma_m^T = \gamma_m^*$. Moreover, $B^* = B^T = -B$, $B^{-1} = B$. The projections on the susy spinor lead to
\begin{equation}
  \epsilon_-^1 = (\lambda_1, 0, 0, 0)^T\,, \qquad
  \epsilon_-^2 = (0, 0, 0, \lambda_2)^T\,.
\end{equation}
There is a symplectic Majorana condition
\begin{equation}\label{eq:symplectic_Majorana_condition}
  \begin{aligned}
    (\epsilon_+^i)^* &= \imath B \epsilon_{ij} \epsilon_+^j\,, &
    (\epsilon_{+i})^* &= \imath B \epsilon^{ij} \epsilon_{+j}\,, \\
    (\epsilon_-^i)^* &= -\imath B \epsilon_{ij} \epsilon_-^j\,, &
    (\epsilon_{-i})^* &= -\imath B \epsilon^{ij} \epsilon_{-j}\,, \\
    (\eta_-^i)^* &= \imath B \epsilon_{ij} \eta_-^j\,, &
    (\eta_{-i})^* &= \imath B \epsilon^{ij} \eta_{-j}\,, \\
    (\eta_+^i)^* &= -\imath B \epsilon_{ij} \eta_+^j\,, &
    (\eta_{+i})^* &= -\imath B \epsilon^{ij} \eta_{+j},\,, \\
  \end{aligned}
\end{equation}
from which it follows that $\lambda_2^* = \lambda_1$. We choose a normalisation such that
\begin{equation}
  \vert \lambda_1 \vert^2 = \vert \lambda_2 \vert^2 = 1\,.
\end{equation}
We could fix the phase, yet there's a danger of deriving phase-dependent expressions. In other words, we choose a normalisation that keeps a phase $\varphi$:
\begin{equation}\label{spinor_normalisatio_phase}
  \lambda_1 = e^{\imath \varphi}\,, \qquad
  \lambda_2 = e^{-\imath \varphi}\,.
\end{equation}
Note that the symplectic Majorana condition implies that
\begin{equation}
  \begin{aligned}
    (\epsilon_+^i)^\dagger &= -\imath (\epsilon_+^j)^T \epsilon_{ji} B\,, &
    (\epsilon_{+i})^\dagger &= -\imath (\epsilon_{+j})^T \epsilon_{ji} B \,,\\
    (\epsilon_-^i)^\dagger &= \imath (\epsilon_-^j)^T \epsilon_{ji} B\,, &
    (\epsilon_{-i})^\dagger &= \imath (\epsilon_{-j})^T \epsilon_{ji} B\,.
  \end{aligned}
\end{equation}
As with the symplectic Majorana condition, the signs for $(\eta_\pm^i)^\dagger$ are switched.

Turning to bilinears, our normalisation corresponds to
\begin{equation}
  \epsilon_-^i B \epsilon_-^j = \imath \epsilon^{ij}\,, \qquad
  (\epsilon_-^i)^\dagger \epsilon_-^j = \delta_i^j\,.
\end{equation}
Due to the chirality, $\epsilon_-^i B \gamma_{(1)} \epsilon_-^j = \epsilon_-^i B \gamma_{(3)} \epsilon_-^j = 0$. However, there are two-forms. We define
\begin{equation}\label{eq:M-triplets}
  \begin{aligned}
    \mathcal{M}_+^{ij} &= \imath \epsilon_+^i B \gamma_{(2)} \epsilon_+^j = V^2
    \begin{pmatrix}
      \lambda_1^2 (e^1 - \imath e^2) \wedge (e^3 + \imath e^4) &
      \imath (e^{12} - e^{34}) \\
      \imath (e^{12} - e^{34}) &
      (\lambda_1^*)^2 (e^1 + \imath e^2) \wedge (e^3 - \imath e^4)
    \end{pmatrix}\, , \\
    \mathcal{M}_-^{ij} &= -\imath \epsilon_-^i B \gamma_{(2)} \epsilon_-^j =
    \begin{pmatrix}
      \lambda_1^2 \Omega & -\imath \omega \\
      -\imath \omega & (\lambda_1^*)^2 \bar{\Omega}
    \end{pmatrix}\, ,
  \end{aligned}
\end{equation}
with $\omega = J = e^{12} + e^{34}$ and $\Omega = (e^1 + \imath e^2) \wedge (e^3 + \imath e^4)$. The factors of $\imath$ render the $SU(2)$ transformations sensible. That is, $(\mathcal{M}_\pm^{ij})^* = \epsilon^{ik} \epsilon^{jl} \mathcal{M}_{\pm kl}$. Moreover,
\begin{equation}\label{eq:M-squared_identity}
  \begin{aligned}
    \mathcal{M}^{ij}_{-mn} [(\mathcal{M}_-^{kl})^*]^{mn} &= 4 (\delta^i_k \delta^j_l + \delta^i_l \delta^j_k)\,,  \\
    \mathcal{M}^{ij}_{+mn} [(\mathcal{M}_+^{kl})^*]^{mn} &= 4 V^4 (\delta^i_k \delta^j_l + \delta^i_l \delta^j_k)\,.
  \end{aligned}
\end{equation}
Similarly, we define
\begin{equation}\label{eq:N-quadruplet}
  \begin{aligned}
    \mathcal{N}^{ij} &= \epsilon_-^i B \gamma_{(1)} \epsilon_+^j\,, \qquad
    \epsilon_+^i B \gamma_{(1)} \epsilon_-^j = -\mathcal{N}^{ji}\,.
  \end{aligned}
\end{equation}
The various matrices are related,
\begin{equation}\label{eq:M-N-matrix-relations}
  \mathcal{N}_a^{ij} + \mathcal{M}^{ij}_{-ab} V^b + \epsilon^{ij} V_a = 0\,, \qquad
  \mathcal{N}_a^{ij} - V^{-2} \mathcal{M}^{ij}_{+ab} V^b + \epsilon^{ij} V_a = 0\,.
\end{equation}

Regarding the raising and lowering of $SU(2)$ indices,
\begin{equation}
  \epsilon_{i+} = \epsilon_{ij} \epsilon_+^j\,, \qquad
  \epsilon_+^i = -\epsilon^{ij} \epsilon_{+j}\,, \qquad
  \epsilon_-^i = \epsilon^{ij} \epsilon_{-j}\,, \qquad
  \epsilon_{-i} = -\epsilon_{ij} \epsilon_-^j\,.
\end{equation}
(In the main text this is $\epsilon^i$ of course). As to the various fields, we use the following.
\begin{equation}
  \begin{aligned}
    \Omega^{Ii}_+ &= \epsilon^{ij} \Omega^I_{+j}\,, &\quad
    \Omega^I_{-i} &= \epsilon_{ij} \Omega^{Ij}_-\,, &\quad
    \psi_{+i} &= \epsilon_{ij} \psi^j_+\,, &\quad
    \psi_-^i &= \epsilon^{ij} \psi_{-j}\,, \\
    \chi_{+i} &= \epsilon_{ij} \chi^j_+\,, &\quad
    \chi_-^i &= \epsilon^{ij} \chi_{-j}\,, &\quad
  \end{aligned}
\end{equation}

\section{Calculation of the cohomological complex}\label{sec:appendix_cohomological_susy_technicalities}

We supplement the discussion of the cohomological complex in section \ref{sec:the_cohomological_complex}. To do so, we introduce some additional forms,
\begin{equation}
  \begin{aligned}
    \chi_+ &= \epsilon_{ij} \epsilon^i_+ B \gamma_{(2)} \Omega^j_+\,, &
    \chi &= \epsilon_{ij} \epsilon^i_- B \gamma_{(2)} \Omega^j_-\,, \\
    \Psi_+ &= \frac{1}{2} \epsilon^{ij} \epsilon_{-i} B \gamma_{(1)} \Omega_{+j}\,, &
    \Psi_- &= \frac{1}{2} \epsilon_{ij} \epsilon_+^i B \gamma_{(1)} \Omega_-^j\,, \\
    \eta_+ &= \frac{\imath}{2} \epsilon_{ij} \epsilon_+^i B \Omega_+^j\,, &
    \eta &= -\frac{\imath}{2} \epsilon_{ij} \epsilon_{-}^i B \Omega_{-}^j\,.
  \end{aligned}
\end{equation}
Of course, $\Psi = \Psi_+ + \Psi_-$. The above are related via
\begin{equation}
  \begin{aligned}
    \eta_+ &= -\imath_{V} \Psi_+\,, &
    V^2 \eta &= - \imath_{V} \Psi_-\,, \\
    V^2 \Psi_+ &= -\frac{\imath}{2} \imath_{V} \chi_+ + \eta_+ V\,, &
    \Psi_- &= \frac{\imath}{2} \imath_{V} \chi - \eta V\,, \\
    \frac{\imath}{4} \chi_+ &= ( \Psi_+ \wedge V )^+\,, &
    -\frac{\imath}{4} V^2 \chi &= ( \Psi_- \wedge V )^-\,.
  \end{aligned}
\end{equation}
The gauginos are recovered from
\begin{equation}
  \begin{aligned}
    \Omega_+^i &= -\imath \left( \Psi_a + \eta V_a - \frac{\imath}{2} \imath_V \chi_a \right) \gamma^a \epsilon_-^i\,, \\
    \Omega_-^i &= \left( \frac{\imath}{8} \chi^{ab} \gamma_{ab} + \eta \right) \epsilon_-^i\,.
  \end{aligned}
\end{equation}

By direct calculation, one finds
\begin{subequations}\label{eq:cohomological_susy_variations_appendix}
\begin{IEEEeqnarray}{rCl}
    \delta \mathcal{A} &=& \Psi\,, \\
    \delta \phi &=& \imath_V \Psi + V^2 \eta\,, \\
    \delta \bar{\phi} &=& \eta\,, \\
    \delta \eta &=& D_{V} \bar{\phi} - g[\phi, \bar{\phi}]\,, \\
    \delta \Psi &=& \imath_V \mathcal{F} + D\phi - V^2 D\bar{\phi} = \imath_V F + D(\phi - V^2 \bar{\phi}) \,, \\
    \delta \chi &=& H\,, \\
    \delta H &=& \left( \pounds_V \chi_- - g [\phi - V^2 \bar{\phi} - \imath_V \mathcal{A}, \chi] \right)^-\,.
\end{IEEEeqnarray}
\end{subequations}
When calculating $\delta H$, the most involves step is the evaluation of $\mathcal{M}_-^{ij} \delta Y_{ij}$. Here,
\begin{equation}
  \mathcal{M}_-^{ij} \delta Y_{ij} = \mathcal{M}_-^{ij} \epsilon_{i k} \epsilon_{j l} (- \epsilon_-^k B \slashed{D} \Omega_+^l + \epsilon_+^k B \slashed{D} \Omega_-^l - 2\imath g \epsilon_-^k B [\phi, \Omega_-^l] + 2\imath g \epsilon_+^k B [\bar{\phi}, \Omega_+^l])\,.
\end{equation}
The last two terms are pretty straightforward. Using
\begin{equation}\label{eq:Mij_gaugino_contractions}
  \begin{aligned}
    \mathcal{M}_-^{ij} \epsilon_{ik} \epsilon_{jl} \epsilon_-^l B \Omega_-^k &= -\vert \lambda_1 \vert^2 \chi\,, \\
    \mathcal{M}_-^{ij} \epsilon_{ik} \epsilon_{jl} \epsilon_+^l B \Omega_+^k &= -4 \imath \vert \lambda_1 \vert^2 (\Psi_+ \wedge V)^-\,,    
  \end{aligned}
\end{equation}
one can rewrite the right hand side as
\begin{equation}
  \mathcal{M}_-^{ij} \epsilon_{i k} \epsilon_{j l}
  \left(- \epsilon_-^k B \slashed{D} \Omega_+^l + \epsilon_+^k B \slashed{D} \Omega_-^l \right)
  + \vert \lambda_1 \vert^2 \left( 2\imath g [\phi, \chi] + 8 g [\bar{\phi}, (\Psi_+ \wedge V)^-]\right)\,.
\end{equation}
Since covariant derivatives \eqref{eq:covariant-derivatives_defined} include coupling to the $SU(2)$ background, the first terms expands to
\begin{equation}
  \mathcal{M}_-^{ij} \epsilon_{i k} \epsilon_{j l} \left\{
    \mp \epsilon_\mp^k B \gamma^m \left( \nabla_m \Omega_\pm^l + g [\mathcal{A}_m, \Omega_\pm^l] + \frac{\imath}{2} A_{(\text{Ric})m} \sigma^{3l}_{\phantom{3l}\hat{l}} \Omega_\pm^{\hat{l}} \right)\,.
  \right\}
\end{equation}
For the moment we ignore the terms involving $[\mathcal{A}, \bullet]$. Then one can show that the remainder is equal to
\begin{equation}
  \begin{aligned}
    \vert \lambda_1 \vert^2 \{ &-2 \imath \imath_V [3 \epsilon_{ij} \epsilon_-^i B \gamma_{[mn} \nabla_{l]} \Omega_-^j dx^l \otimes dx^m \otimes dx^n ]\\
        &- 2\imath \imath_V [ 3 \frac{\imath}{2} \epsilon_{ij} \epsilon_-^i B \gamma_{[mn} A_{(\text{Ric})l]} \sigma^{3j}_{\phantom{3j}k} \Omega_-^k ] \\
    &- 4 V \wedge [-\frac{\imath}{2} \epsilon_{ij} \epsilon_-^i B \nabla_m \Omega_-^j dx^m] \\
    &+ 4 [-\epsilon_{ij} \epsilon_-^i B \gamma_{[n} \nabla_{m]} \Omega_+^j dx^m \otimes dx^n ] \\&+ 4 [- \frac{\imath}{2} \epsilon_{ij} \epsilon_-^i B \gamma_{[n} A_{(\text{Ric})m]} \Omega_+^j dx^m \otimes dx^n ]
     \}^- \\
     = \vert \lambda_1 \vert^2 \{ &-2\imath \imath_V d\chi + 4 d\eta \wedge V + 4 d\Psi_+ \}^-.
  \end{aligned}
\end{equation}
The last step uses\footnote{
  For convenience, recall:
  \begin{equation*}
      \nabla_m \epsilon_-^i = -\frac{\imath}{2} A_{(\text{Ric})m} \sigma^{3i}_{\phantom{3i}j} \epsilon_-^j, \qquad
      \nabla_m \epsilon_+^i = -\frac{\imath}{2} A_{(\text{Ric})m} \sigma^{3i}_{\phantom{3i}j} \epsilon_+^j - \frac{\imath}{4} T_{mn} \gamma^n \epsilon_-^i.
  \end{equation*}
}
\begin{equation}\label{eq:dPsi}
  \begin{aligned}
    d\eta_{m} &= -\frac{\imath}{2} \epsilon_{ij} \epsilon_-^i B \left( \nabla_m  \Omega_-^j + \frac{\imath}{2} A_{(\text{Ric})m} \sigma^{3j}_{\phantom{3j}k} \Omega_-^k \right), \\
    d\Psi_{+mn} &= - \epsilon_{ij} \epsilon_-^i B \gamma_{[n} \left(\nabla_{m]} \Omega_+^j + \frac{\imath}{2} A_{(\text{Ric})m]} \sigma^{3j}_{\phantom{3j}k} \Omega_+^k \right), \\
    d\chi_{lmn} &= 3 \epsilon_{ij} \epsilon_-^i B \gamma_{[lm} \left( \nabla_{k]} \Omega_-^j + \frac{\imath}{2} A_{(\text{Ric})k]} \sigma^{3j}_{\phantom{3j}k} \Omega_-^k \right)\,,
  \end{aligned}
\end{equation}
which follow directly from the definitions of the forms.\footnote{
One can derive a similar equation for $\Psi_-$:
\begin{equation*}
      d\Psi_{-mn} = \epsilon_{ij} \epsilon_+^i B \gamma_{[n} \left(\nabla_{m]} \Omega_-^j + \frac{\imath}{2} A_{(\text{Ric})m]} \sigma^{3j}_{\phantom{3j}k} \Omega_-^k \right)
    + \frac{1}{2} T_{mn} \eta - \frac{\imath}{4} T_{[m \vert s} \chi^{s}_{\phantom{s}\vert n]}\,.
\end{equation*}
}
For the commutator terms, we use
\begin{equation}
  \begin{aligned}
    \mathcal{M}_-^{ij} \epsilon_{ik} \epsilon_{jl} \epsilon_-^k B \gamma^m [\mathcal{A}_m, \Omega_+^l] &= 4 \vert \lambda_1 \vert^2 [\mathcal{A}, \Psi_+]^-\,, \\
    \mathcal{M}_-^{ij} \epsilon_{ik} \epsilon_{jl} \epsilon_+^k B \gamma^m [\mathcal{A}_m, \Omega_-^l] &= \vert \lambda_1 \vert^2 \left( 4 [\mathcal{A}, \eta V]  - 2\imath \imath_V [\mathcal{A}, \chi] \right)^-\, .
  \end{aligned}
\end{equation}
Returning to $\mathcal{M}_-^{ij} \delta Y_{ij}$,
\begin{equation}\label{eq:SUSY_variation_M-Y}
  \begin{aligned}
    \mathcal{M}_-^{ij} \delta Y_{ij} = \vert \lambda_1 \vert^2 \{ &-2\imath \imath_V d\chi - 2\imath g \imath_V [\mathcal{A}, \chi] + 2\imath g [\phi, \chi] \\
    &+ 4 d\eta \wedge V + 4 g [\mathcal{A}, \eta V] \\
    &+ 4 (d+g[\mathcal{A}, \bullet]) \Psi_+ + 8 g [ \bar{\phi}, \Psi_+ \wedge V] \}^-\, .
  \end{aligned}
\end{equation}
Of course, $\imath_V [\mathcal{A}, \chi] = [\imath_V \mathcal{A}, \chi] - [\mathcal{A}, \imath_V \chi]$.

The SUSY transformations of the variables introduced here are as follows (the matrices $\mathcal{N}^{ij}$ are defined in equation \eqref{eq:N-quadruplet}):
\begin{subequations}\label{eq:full_cohomological_susy_variations}
\begin{IEEEeqnarray}{rCl}
    \delta \mathcal{A} &=& \Psi_+ + \Psi_-\,, \\
    \delta \phi &=& \eta_+\,, \\
    \delta \bar{\phi} &=& \eta\,, \\
    \delta \eta_+ &=& D_{V} \phi - V^2 [\phi, \bar{\phi}]\,, \\
    \delta \eta &=& D_{V} \bar{\phi} - [\phi, \bar{\phi}]\,, \\
    \delta \Psi_+ &=& D \phi - g [\phi, \bar{\phi}] V + \frac{1}{4} \mathcal{N}^{ij} Y_{ij} + \imath_{V} \mathcal{F}^+\,, \\
    \delta \Psi_- &=& - V^2 D\bar{\phi} + g [\phi, \bar{\phi}] V - \frac{1}{4} \mathcal{N}^{ij} Y_{ij} + \imath_{V} \mathcal{F}^-\,, \\
    \delta \Psi &=& \imath_V \mathcal{F} + D\phi - V^2 D\bar{\phi} = \imath_V F + D(\phi - V^2 \bar{\phi}) \,, \\
    \delta \chi &=& - 2 \imath \mathcal{F}^- + \frac{\imath}{2} \mathcal{M}^{ij}_- Y_{ij} - 4 \imath (D\bar{\phi} \wedge V)^- = H\,, \\
    \delta \chi_+ &=& -2 \imath V^2 \mathcal{F}^+ + \frac{\imath}{2} \mathcal{M}^{ij}_+ Y_{ij} + 4 \imath (D\phi \wedge V)^+\,, \\
    \mathcal{M}_-^{ij} \delta Y_{ij} &=& \vert \lambda_1 \vert^2 \{ 2\imath (- \imath_V d\chi - g \imath_V [\mathcal{A}, \chi] + g [\phi, \chi] ) \nonumber \\
    &&+ 4 D \eta \wedge V
    + 4 d_\mathcal{A} \Psi_+ + 8 g [ \bar{\phi}, \Psi_+ \wedge V] \}^-\,, \\
    \delta \mathcal{F}^- &=& (d_\mathcal{A}\Psi)^- - \frac{1}{2} \eta T^- \nonumber\\ 
    &=& (d_\mathcal{A}\Psi_+)^- + \frac{\imath}{2} ( d\imath_V \chi + g [\mathcal{A}, \imath_V \chi ] )^- 
    - (D\eta \wedge V)^-\,,\\
    \delta (D\bar{\phi} \wedge V)^- &=& (D\eta \wedge V)^- 
    + g [(\Psi_+ \wedge V)^-, \bar{\phi}] - \frac{\imath}{4} g V^2 [\chi, \bar{\phi}], \\
    \delta H &=& \left( \pounds_V \chi - g [\phi - V^2 \bar{\phi} - \imath_V \mathcal{A}, \chi] \right)^-\,.
\end{IEEEeqnarray}
\end{subequations}
Here we used the notation
\begin{equation}
  \begin{aligned}
    d_\mathcal{A}\Psi &= (D_m \Psi_n - D_n \Psi_m) dx^m \otimes dx^n = d \Psi + g[\mathcal{A}, \Psi]\,,
  \end{aligned}
\end{equation}
with $D_m \Psi_n$ as in \eqref{eq:Yang-Mills_conventions}.

There is another minor technicality involving $\delta \Psi = \delta^2 \mathcal{A}$. The supersymmetry variations yield
\begin{equation}
  \delta^2 \mathcal{A} = \imath_V \mathcal{F} + (D\phi - V^2 D\bar{\phi})\,.
\end{equation}
Using
\begin{equation}
  V^2 D\bar{\phi} = D( V^2 \bar{\phi} ) - \frac{1}{4} \imath_V T \bar{\phi}\,,
\end{equation}
however, one can show that
\begin{equation}
  \delta^2 \mathcal{A} = \imath_{V} F + D(\phi - V^2 \bar{\phi})\,,
\end{equation}
where now the actual field strength $F$ makes an appearance instead of $\mathcal{F}$. With
$\imath_V F = \pounds_V \mathcal{A} - D (\imath_V \mathcal{A})$ it follows that
\begin{equation}
  \delta^2 \mathcal{A} = \pounds_{V} \mathcal{A} + D( \phi - V^2 \bar{\phi} - \imath_{V} \mathcal{A})\,.
\end{equation}

\section{The localization locus}
\label{sec:appendix_localization_locus}

Turning to the gauge field, we adapt the argument of \cite{Gomis:2011pf} to the case at hand. The relevant equations are
\begin{equation}
  \begin{aligned}
    0 &= \imath_V (F + \phi_1 dV) + (1-V^2) D\phi_1,\, \\
    0 &= (F + \phi_1 dV)^- + 2 (D\phi_1 \wedge V)^-\,.
  \end{aligned}
\end{equation}
Combining the two,
\begin{equation}
  \imath_V \star (F + \phi_1 dV) = - (1+V^2) D\phi_1\,.
\end{equation}
Integrating with a yet undetermined measure $\mu$, (note the sign due to \eqref{eq:hodge_inner_product_sign})
\begin{equation}
  \begin{aligned}
    0 &= -\int \mu \left[ \imath_V \star (F + \phi_1 dV) + (1+V^2) D\phi_1 \right] \wedge \star \left[ \imath_V \star (F + \phi_1 dV) + (1+V^2) D\phi_1 \right] \\
    &= \int \mu \left\vert \imath_V \star (F + \phi_1 dV) \right\vert^2 + \mu \left\vert (1+V^2) D\phi_1 \right\vert^2 - 2\mu (1+V^2) D\phi \wedge \star \imath_V \star (F + \phi_1 dV)\,.
  \end{aligned}
\end{equation}
We focus on the cross term. Up to an overall sign and factor that play no importance, this is
\begin{equation}
  \mu (1+V^2) D\phi_1 \wedge (F + \phi_1 dV) \wedge V\,.
\end{equation}
Setting $\hat{A} = A + \phi_1 V$ we have $\hat{F} = F + \phi_1 dV + D\phi \wedge V$ and $\hat{D} \phi_1 = D \phi_1$; also $\imath_V \hat{F} = - D\phi_1$. Moreover, we choose $\mu = V^{-2} (1 + V^2)^{-1}$. Then
\begin{equation}
  D\phi_1 \wedge \hat{F} \wedge (V^{-2} V) = \hat{D} (\phi_1 \hat{F}) \wedge (V^{-2} V) = \hat{D} (\phi_1 \hat{F} \wedge V^{-2} V) - \phi_1 \hat{F} \wedge d(V^{-2} V)\,.
\end{equation}
At this point, we drop the total derivative. Focussing on the remainder, we note that
\begin{equation}
  d(V^{-2} V) = V^{-2} dV + V^{-4} \imath_V dV \wedge V = \imath_V (V^{-4} dV \wedge V)\,.
\end{equation}
Therefore
\begin{equation}
  \phi_1 \hat{F} \wedge d(V^{-2} V) = \phi_1 \hat{F} \imath_V (V^{-4} dV \wedge V) = - V^{-4} \phi_1 \imath_V \hat{F} \wedge dV \wedge V\,.
\end{equation}
Substituting our earlier result $\imath_V \hat{F} = - D\phi_1$,
\begin{equation}
  \begin{aligned}
    \dots &= \frac{1}{2} V^{-4} \hat{D} (\phi^2) \wedge dV \wedge V \\
    &= \frac{1}{2} \hat{D} (\phi^2 V^{-4} dV \wedge V) - \frac{1}{2} \phi^2 dV \wedge d(V^{-4} V) \\
    &= \frac{1}{2} \hat{D} (\phi^2 V^{-4} dV \wedge V) - \frac{1}{2} \phi^2 V^{-6} (V^2 dV \wedge dV + 2 dV \wedge \imath_V dV \wedge V)\,.
  \end{aligned}
\end{equation}
Again discarding the total derivative, we note that
\begin{equation}
  \imath_V (V^2 dV \wedge dV + 2 dV \wedge \imath_V dV \wedge V) = 0\,,
\end{equation}
from which it follows that the term in parentheses vanish. Thus the cross term vanishes and we are left with
\begin{equation}
    0 = \int_{M_4} \frac{1}{V^2 (1+V^2)} \left[ \left\vert \imath_V \star (F + \phi_1 dV) \right\vert^2 + \left\vert (1+V^2) D\phi_1 \right\vert^2 \right]\,.
\end{equation}
Clearly
\begin{equation}
  D \phi_1 = 0\,,
\end{equation}
and substituting this in our previous equations,
\begin{equation}
  0 = \imath_V (F + \phi_1 dV) = (F + \phi_1 dV)^- = \imath_V \star (F + \phi_1 dV)\,.
\end{equation}
Note that for a generic two-form $\omega$,
\begin{equation}
  V^2 \omega = \imath_V (V \wedge \omega) + V \wedge (\imath_V \omega) = \{ \imath_V, V \wedge \} \omega\,.
\end{equation}
In other words, knowing $\imath_V \omega$ and $V \wedge \omega$ is enough to reconstruct the form. As we remarked earlier
\begin{equation}
  \star \imath_V \star (F + \phi_1 dV) = \pm (F + \phi_1 dV) \wedge V\,,
\end{equation}
so we can conclude that
\begin{equation}
  F + \phi_1 dV = 0\,.
\end{equation}
In other words,
\begin{equation}
  \mathcal{A} = - \phi_1 V, \qquad D \phi_1 = d \phi_1 = 0\,.
\end{equation}

\section{The tree level action}\label{sec:appendix_tree_level}

We complement the discussion of the tree level action in section \ref{sec:tree_level_contribution}. At the localization locus given by \eqref{eq:BPS_solution} and \eqref{eq:ghost_sector_localization_locus}, the non-vanishing terms contributing to \eqref{eq:vector_multiplet_action} are
\begin{equation}
  \begin{aligned}
    \mathcal{M}^{ij}_- Y_{ij} &= -4\imath (\phi_2 dV + 2 D\phi_2 \wedge V)^-\,, \\
    (d\phi_2)^2 &= (a_0^E)^2 \left( \frac{\imath_V dV}{(1+V^2)^2} \right)^2\,, \\
    \phi^2 (T^+)^2 &= - 4 (a_0^E)^2 \left( \frac{dV^+}{1+V^2} \right)^2\,.
  \end{aligned}
\end{equation}
We also need
\begin{equation}
  \frac{1}{8} Y^i_{\phantom{i}j} Y^j_{\phantom{j}i} = \frac{1}{64} (\mathcal{M}^{ij}_{-} Y_{ij})^{mn} (\mathcal{M}^{kl}_- Y_{kl})^*_{mn}\,,
\end{equation}
which follows from \eqref{eq:M-squared_identity}. These combine to
\begin{equation}\label{eq:tree_level_action_substituted}
  (a_0^E)^2 \int_{M_4} \frac{\vol}{(1+V^2)^2} \left\{ \frac{1}{4} \left[ \frac{2(\imath_V dV \wedge V)^-}{1+V^2} + dV^- \right]^2 + \frac{1}{(1+V^2)^2} (\imath_V dV)^2 + \frac{1}{4} (dV^+)^2 \right\}\,.
\end{equation}
To proceed, we study this term by term.

First we note that
\begin{equation}
  \begin{aligned}
    &\left( \frac{2 \imath_V dV \wedge V^-}{1+V^2} + dV^- \right)^2 \\
    = &\frac{4}{(1+V^2)^2} (\imath_V dV \wedge V^-)^2 + (dV^-)^2 + \frac{4}{1+V^2} (\imath_V dV \wedge V)^-_{mn} dV^{-mn}\,.
  \end{aligned}
\end{equation}
The first of these three terms can be rewritten using
\begin{equation}
  \begin{aligned}
    \vol (\imath_V dV \wedge V^-)^2 &= \vol \left[ \frac{1}{2} (\imath_V dV \wedge V)^2 - \frac{1}{2} (\imath_V dV \wedge V)_{mn} \star (\imath_V dV \wedge V)^{mn} \right] \\
    &= \vol V^2 (\imath_V dV)^2 + \imath_V dV \wedge V \wedge \imath_V dV \wedge V = \vol V^2 (\imath_V dV)^2\,.
  \end{aligned}
\end{equation}
The crossterm on the other hand is
\begin{equation}
  \begin{aligned}
    \vol (\imath_V dV \wedge V^-)_{mn} dV^{-mn} &= -2 \imath_V dV \wedge V \wedge \star dV^- \\
    &= - \imath_V dV \wedge V \wedge dV + \imath_V dV \wedge V \wedge \star dV \\
    &= -\frac{1}{2} \imath_V (dV \wedge dV) \wedge V + dV \wedge V \wedge \star (dV \wedge V) \\
    &=-\frac{1}{2} V^2 dV \wedge dV - \vol (\imath_V dV)^2\,.
  \end{aligned}
\end{equation}
Therefore
\begin{equation}
  \begin{aligned}
    \vol \left( \frac{2 \imath_V dV \wedge V^-}{1+V^2} + dV^- \right)^2 &= \vol \left[ -\frac{4}{(1+V^2)^2} (\imath_V dV)^2 + (dV^-)^2 \right] - \frac{2V^2}{1+V^2} dV \wedge dV\,.    
  \end{aligned}
\end{equation}

Returning to \eqref{eq:tree_level_action_substituted}, the $(\imath_V dV)^2$ terms cancel and the expression simplifies to
\begin{equation}
  (a_0^E)^2 \int_{M_4} \frac{1}{(1+V^2)^2} \left( \frac{\vol}{4} dV^2 - \frac{V^2}{2(1+V^2)} dV \wedge dV \right)\,.
\end{equation}
We can rewrite this as
\begin{equation}
  -\frac{(a_0^E)^2}{2} \int_{M_4} \frac{1}{(1+V^2)^2} \left( dV \wedge \star dV + \frac{V^2}{(1+V^2)} dV \wedge dV \right)\,.
\end{equation}
Here the second term vanishes. To see this, consider a generic function $f(V^2)$. One finds $\int f(V^2) dV \wedge dV = - \frac{1}{2} \int f^\prime(V^2) V^2 dV \wedge dV$. Thus
\begin{equation}
  \int_{M_4} [2 f(V^2) + f^\prime (V^2) V^2] dV \wedge dV = 0\,.
\end{equation}
Observing that
\begin{equation}
  f(V^2) = \frac{1}{V^4} \left[ \frac{3+4V^2}{2(1+V^2)^2} + \log (V^2 + 1) + \text{const} \right]
\end{equation}
satisfies
\begin{equation}
  2 f(V^2) + f^\prime (V^2) V^2 = \frac{V^2}{(1+V^2)^3}
\end{equation}
settles things.

Finally we arrive at the result of this section. The tree level action is given by
\begin{equation}
  -\frac{(a_0^E)^2}{2} \int_{M_4} \frac{1}{(1+V^2)^2} dV \wedge \star dV = \frac{(a_0^E)^2}{4} \int_{M_4} \left( \frac{dV}{1+V^2}\right)^2\,.
\end{equation}

\section{Toric K\"ahler geometry}
\label{sec:appendix_abreu}

Consider a toric K\"ahler manifold $M_4$. The image of the moment map $\phi : M_4 \to \bR^4$ is the Delzant polytope $P = \phi (M_4) \subset \bR^4$, $P^\circ$ its interior. Then $M_4^\circ = \phi^{-1}(P^\circ)$ is an open dense subset of $M_4$. $M_4^\circ$ is symplectomorphic to $P^\circ \times \mathbb{T}^2$ and we can introduce symplectic coordinates $(x,y)$:
\begin{equation}
  M_4^\circ \cong P^\circ \times \mathbb{T}^2 = \left\{ (x,y) \vert x \in P^\circ \subset \bR^2, y \in \bR^2 / \bZ^2 \right\}\,.
\end{equation}
The torus acts as
\begin{equation}
  t \cdot (x, y) = (x, y+t) \qquad t \in \mathbb{T}^2
\end{equation}
and the symplectic form is $\omega = dx \wedge dy = \sum_i dx^i \wedge dy_i$. The volume form is trivially $\frac{1}{2} \omega \wedge \omega$, and Guillemin \cite{guillemin1994kaehler} tells us that the symplectic volume of $M_4$, $\int_{M_4} e^\omega$ is $(2\pi)^2$, the Euclidean volume of the Delzant polytope $\Delta$.

The work of Guillemin and Abreu \cite{guillemin1994kaehler,abreu2003kahler} gives us a simple recipe to calculate the \emph{canonical almost complex structure} and metric from the Delzant polytope. The Delzant polytope can be described by a set of inequalities of the form $\langle x, \mu_r \rangle \geq \lambda_r$, $r = 1, \dots, d$, each $\mu_r$ being a primitive element of the lattice $\bZ^2 \subset \bR^2$ and inward pointing normal to the $r$-th $(n-1)$-dimensional face of $P$. Consider the affine functions $l_r : \bR^2 \to \bR$, defined by
\begin{equation}
  l_r (x) = \langle x, \mu_r \rangle - \lambda_r\,.
\end{equation}
The function
\begin{equation}
  g_P(x) = \frac{1}{2} \sum_{r=1}^d l_r(x) \log l_r(x)
\end{equation}
is smooth on $P^\circ$. Define $G_P = \text{Hess}_x (g)$, i.e.~$(G_P)_{ij} = \partial_{x^i} \partial_{x^j} g$. Then
\begin{equation}
  J_P = \begin{pmatrix}
    0 & - G_P^{-1} \\ G_P & 0
  \end{pmatrix},
  \qquad
  ds^2 = \begin{pmatrix}
    G_P & 0 \\ 0 & G_P^{-1}
  \end{pmatrix}.
\end{equation}
We can calculate the Christoffel symbols
\begin{equation}
  \begin{aligned}
    \Gamma^{y_i}_{y_k x^l} &= - \frac{1}{2} G_{ij} \partial_l (G^{-1})^{jk}\,, &
    \Gamma^{x^i}_{y_k y_l} &= \frac{1}{2} (G^{-1})^{ij} \partial_j (G^{-1})^{kl}\,, &
    \Gamma^{x^i}_{x^k x^l} &= \frac{1}{2} (G^{-1})^{ij} \partial_k G_{jl}\,,
  \end{aligned}
\end{equation}
as well as the Ricci tensor.

Now, we pick a generic Killing vector $V = V[p,q] = p \partial_{y_1} + q \partial_{y_2}$; $V = V_i \partial_{y_i}$. It's norm is
\begin{equation}
  V^2 = p^2 (G_P^{-1})^{11} + q^2 (G_P^{-1})^{22} + 2 pq (G_P^{-1})^{12}\,.
\end{equation}
And of course $G_P^{-1} = (\det G_P)^{-1} G^\# = (\det G_P)^{-1} \left( \begin{smallmatrix} G_{22} & -G_{12} \\ -G_{12} & G_{11} \end{smallmatrix}\right)$. There is a dual form $v = v[p,q]$ which satisfies ($\epsilon^1_{\phantom{1}2} = 1 = \epsilon_1^{\phantom{1}2}$)
\begin{equation}
  \begin{aligned}
    v &= \langle V, \cdot \rangle = \left( p (G_P^{-1})^{11} + q (G_P^{-1})^{12} \right) dy_1 + \left( p (G_P^{-1})^{21} + q (G_P^{-1})^{22} \right) dy_2\,, \\
    dv &= \partial_{x^i} v_{y_j} dx^i \wedge dy_j\,, \\
    \star dv &= -\partial_{x^i} v_{y_j} \epsilon^i_{\phantom{i}k} \epsilon_j^{\phantom{j}l} dx^k \wedge dy_l, \\
    dv^\pm &= \frac{1}{2} (\delta^i_k \delta^j_l \mp \epsilon^i_{\phantom{i}k} \epsilon_j^{\phantom{j}l}) \partial_{x^i} v_{y_j} dx^k \wedge dy_l\,.
  \end{aligned}
\end{equation}
With $\langle dx^i \wedge dy_j, dx^k \wedge dy_l \rangle = 2 (G_P^{-1})^{ik} (G_P)_{jl}$,
\begin{equation}
  \begin{aligned}
    (dv)^2 &= 2 \partial_{x^i} v_{y_j} \partial_{x^k} v_{y_l} (G_P^{-1})^{ik} (G_P)_{jl}\, , \\
    (dv^\pm)^2 &= \frac{1}{2} \left( \delta^i_k \delta^l_j \mp \epsilon^i_{\phantom{i}k} \epsilon_j^{\phantom{j}l} \right) \left( \delta^{\hat{i}}_{\hat{k}} \delta^{\hat{l}}_{\hat{j}} \mp \epsilon^{\hat{i}}_{\phantom{\hat{i}}\hat{k}} \epsilon_{\hat{j}}^{\phantom{\hat{j}}\hat{l}} \right) (\partial_{x^i} v_{y_j}) (\partial_{x^{\hat{i}}} v_{y_{\hat{j}}}) (G_P^{-1})^{k\hat{k}} (G_P)_{l\hat{l}}\,.
  \end{aligned}
\end{equation}
One can massage these terms a bit as $\partial_{x^i} G^{-1} = -G^{-1} \partial_{x^i} G G^{-1}$.
\begin{equation}
  \begin{aligned}
    dv &= -(p \; q)_i (G_P^{-1})^{ij} (G_P^{-1})^{kl} (\partial_m \partial_j \partial_k g_P) dx^m \wedge dy_l\,.
  \end{aligned}
\end{equation}

\subsection*{Local form of the $\CP^2$ metric}

For $\CP^2$, we have
\begin{equation}
  l_1 = x_1\,, \qquad
  l_2 = x_2\,, \qquad
  l_3 = 1-x_1-x_2\,.
\end{equation}
Therefore
\begin{equation}
  G_P^{-1} = 2
    \begin{pmatrix}
      x_1 (1-x_1) & -x_1 x_2 \\ -x_1 x_2 & x_2 (1-x_2)
    \end{pmatrix}\,.
\end{equation}
The polytope is the triangle bounded by $x_1 = 0$, $x_2 = 0$, and $x_1 + x_2 = 1$.

Near $(0,0)$, the metric takes the form
\begin{equation}
 ds^2 = \frac{1}{2 x_1} dx_1^2 + 2 x_1 dy_1^2 + \frac{1}{2 x_2} dx_2^2 + 2 x_2 dy_2^2\,, \qquad
 V = \sum_i p_i \partial_{y_i}\,.
\end{equation}
Introducing $2 x_i = \rho_i^2$ brings this in polar coordinates, $ds^2 = \sum_i d\rho_i^2 + \rho_i^2 dy_i^2$. Subsequently we can introduce cartesian coordinates $X_i = \rho_i \cos y_i$, $Y_i = \rho_i \sin y_i$. Then
\begin{equation}
  ds^2 = \sum_i dX_i^2 + dY_i^2\,, \qquad
  V = \sum_i p_i ( - Y_i \partial_{X_i} + X_i \partial_{Y_i} )\,.
\end{equation}
For the Killing spinors this means that
\begin{equation}
  \epsilon_+^i = - \imath p_1 (Y_1 \gamma_1 - X_1 \gamma_2) \epsilon_-^i - \imath p_2 (Y_2 \gamma_3 - X_2 \gamma_4) \epsilon_-^i\,.
\end{equation}
Thus, around $(0,0)$ we have $\Omega$-parameters $(p_1, p_2)$.

Next we study the metric, vector and spinors around $(1,0)$. Here,
\begin{equation}
  ds^2 = \frac{1}{2(1-x_1 - x_2)} (dx_1 + dx_2)^2 + 2 (1-x_1 - x_2) dy_1^2 + \frac{1}{2x_2} dx_2^2 + 2 x_2 (dy_1 - dy_2)^2\,.
\end{equation}
If we introduce coordinates $\hat{x}_1 = 1-x_1-x_2$, $\hat{x}_2 = x_2$, $\hat{y}_1 = y_1$, $\hat{y}_2 = y_1 - y_2$, the metric takes the form of the previous case and
\begin{equation}
  V = p_1 \partial_{\hat{y}_1} + (p_1 - p_2) \partial_{\hat{y}_2}\,.
\end{equation}
By comparison with the previous case, the $\Omega$-parameters are $(p_2-p_1,-p_1)$.

Similarly, the metric around $(0,1)$ is
\begin{equation}
  ds^2 = \frac{1}{2 x_1} dx_1^2 + 2 x_1 (dy_1 - dy_2)^2 + \frac{1}{2(1-x_1-x_2)} (dx_1+dx_2)^2 + 2 (1-x_1-x_2) dy_2^2\,,
\end{equation}
and we perform a coordinate transformations to $\hat{x}_1 = x_1$, $\hat{x}_2 = 1 - x_1 - x_2$, $\hat{y}_1 = y_1 - y_2$, and $\hat{y}_2 = y_2$. Then
\begin{equation}
  V = (p_1 - p_2) \partial_{\hat{y}_1} + p_2 \partial_{\hat{y}_2}\,.
\end{equation}
The epsilon parameters are thus $(-p_2,\,p_1-p_2)$.

The same result can be obtained as well starting with the metric in \eqref{eq:explicitmetric} and graviphoton in \eqref{eq:graviphoton}. In these coordinates the set $V^2=0$ is $\rho=0$, $\{\rho=\frac{\pi}{2},\,\theta=0\}$ and $\{\rho=\frac{\pi}{2},\,\theta=\pi\}$. Expanding the background around each of these points one can see that, around them, in the appropriate coordinates it becomes a copy of flat space  $ds^2=dx_i^2$ such that

\begin{equation}
 V=\epsilon^{(i)}_1\,(x_1\,\partial_{x_2}-x_2\,\partial_{x_1})+\epsilon^{(i)}_2\,(x_3\,\partial_{x_4}-x_4\,\partial_{x_3})\, ,\quad T^+=-2\,(\epsilon^{(i)}_1-\epsilon^{(i)}_2)\,(dx_1\wedge dx_2-dx_3\wedge dx_4)\, ;
\end{equation}
being $\epsilon_{1,\,2}^{(i)}$ given by \eqref{eq:epsilons}.

\bibliographystyle{ytphys}
\small\baselineskip=.97\baselineskip
\bibliography{ref}

\end{document}